\documentclass[12pt]{article}
\pdfoutput=1
\usepackage[nosort]{cite}
\usepackage[height=8.85in,width=6.45in]{geometry}
\usepackage{pifont}
\newcommand{\cmark}{\ding{51}}
\newcommand{\xmark}{\ding{55}}

\usepackage{times}

\usepackage[utf8]{inputenc}
\usepackage{amsmath}
\usepackage{amssymb}
\usepackage{mathtools}
\numberwithin{equation}{section}
\usepackage{slashed}
\usepackage{braket}
\usepackage[svgnames,psnames]{xcolor}
\usepackage[colorlinks,citecolor=DarkGreen,linkcolor=FireBrick,linktocpage,pagebackref]{hyperref}
\usepackage{cite}
\usepackage{graphicx}
\usepackage{tikz}
\usepackage{tikz-cd}

\usepackage{courier}
\usepackage{bm}

\usepackage{dashbox}
\usepackage{subcaption}
\usepackage{enumitem}

\usepackage{mdframed}

\usepackage{blkarray}
\usepackage{arydshln}
\usepackage{dsfont}
\usetikzlibrary{patterns}
\usetikzlibrary{arrows.meta}

\usepackage{calc}

\usepackage{accents}

\newcommand{\ie}{\begin{equation}\begin{aligned}}
\newcommand{\fe}{\end{aligned}\end{equation}}

\renewcommand{\title}[1]{\vbox{\center\LARGE{#1}}\vspace{5mm}}
\renewcommand{\author}[1]{\vbox{\center#1}\vspace{5mm}}
\newcommand{\address}[1]{\vbox{\center\em#1}}

\begin{document}
 
\begin{titlepage}
 	\hfill  	YITP-SB-2022-21,	MIT-CTP/5433
 	\\

\title{Non-invertible Global Symmetries\\ in the Standard Model}

\author{Yichul Choi${}^{1,2}$,   Ho Tat Lam${}^3$, and Shu-Heng Shao${}^1$}

		\address{${}^{1}$C.\ N.\ Yang Institute for Theoretical Physics, Stony Brook University\\
        ${}^{2}$Simons Center for Geometry and Physics, Stony Brook University\\
		${}^{3}$Center for Theoretical Physics, Massachusetts Institute of Technology
		}

\abstract

We identify infinitely many non-invertible generalized   global symmetries in QED and QCD for the real world in the massless  limit.  
In QED, while there is no conserved Noether current for the $U(1)_\text{A}$ axial symmetry because of the ABJ anomaly, for every rational angle $2\pi p/N$, we construct a conserved and gauge-invariant topological symmetry operator. 
Intuitively,  it is a composition of the axial rotation and a fractional quantum Hall state coupled to the electromagnetic $U(1)$ gauge field.  
These conserved symmetry operators do not obey a group multiplication law, but a non-invertible fusion algebra over  TQFT coefficients. 
They act invertibly on all local operators as axial rotations, but non-invertibly on the 't Hooft lines. 
These non-invertible symmetries lead to selection rules, which are consistent with the scattering amplitudes in QED. 
We further generalize our construction to QCD, and show that   the coupling $\pi^0 F\wedge F$ in the  effective pion Lagrangian is necessary to match these non-invertible symmetries in the UV.  
Therefore, the conventional argument for the neutral pion decay  using the ABJ anomaly is now rephrased as a matching condition of a generalized global symmetry.

\end{titlepage}

\eject

\tableofcontents

\section{Introduction}
 
Global symmetry is one of the few intrinsic characteristics of a quantum system that  is invariantly matched across all different descriptions and dualities.  
The most familiar example of a global symmetry is a $U(1)$ global symmetry with a conserved Noether current $j_\mu(x)$. 
Thanks to the conservation equation, $\partial^\mu j_\mu = 0$, the charge $Q= \int d^3x j_0$ is conserved under time evolution, and so is the symmetry operator $U_\vartheta=\exp(i \vartheta Q)$ labeled by a $U(1)$ group element $\vartheta \in [0,2\pi)$. 
In relativistic quantum field theory (QFT), time is on equal footing as any other direction in spacetime. 
We can therefore define the symmetry operator on a general closed three-manifold $M$ in the 3+1-dimensional spacetime
\ie\label{introU}
U_\vartheta(M) = \exp\left( i \vartheta\oint_M \star j\right)\,,
\fe
where $\star$ is the Hodge dual of a differential form.  
When $M$ is a spatial slice, $U_\vartheta(M)$ is an operator acting on the Hilbert space. 
When $M$ extends in the time direction, $U_\vartheta(M)$ is  a defect that changes the boundary condition.\footnote{Since we focus on relativistic QFT in this paper, we will often use the terms ``operator" and ``defect" interchangeably.} 
In this relativistic setting, the conservation under time evolution is upgraded to the statement that   $U_\vartheta(M)$  is a \textit{topological} operator that depends on the choice of the three-manifold $M$ only topologically \cite{Gaiotto:2014kfa}.  
In the case of a $U(1)$ symmetry, the topological nature simply follows from the divergence theorem.

Given a quantum system with a global symmetry, one can attempt to gauge the symmetry to obtain a different system. 
This is, however, not always possible. 
The obstruction is sometimes referred to as the 't Hooft anomaly of a global symmetry, which leads to  nontrivial 't Hooft anomaly matching conditions constraining the  renormalization group flow \cite{tHooft:1979rat}.  
A global symmetry with an 't Hooft anomaly is otherwise completely healthy; one simply cannot gauge  it.

In contrast, the Adler-Bell-Jackiw (ABJ) anomaly \cite{PhysRev.177.2426,Bell:1969ts} (see  \cite{tHooft:1986ooh} for  a review) is the statement that a classical global symmetry fails to persist at the quantum level.  
The ABJ anomaly has many important  phenomenological consequences, including the determination of the neutral pion decay coupling $\pi^0 F\wedge F$ in the effective pion Lagrangian.  
However, if the punchline of the ABJ anomaly is the absence of a global symmetry, why does it imply anything nontrivial in the IR effective Lagrangian?  
More generally, is it meaningful to discuss the ABJ anomaly in a  QFT without a Lagrangian description in terms of a fermion path integral? 
In this paper, for ABJ anomalies where all the participating symmetries are $U(1)$, we will reinterpret them  in terms  of certain generalized global symmetries. 

We start with the ABJ anomaly of the axial $U(1)_\text{A}$ symmetry in the 3+1d massless QED as a warm-up example.  
While there is no gauge-invariant Noether current, for every rational angle $\alpha = 2\pi p/N$, we can dress the naive $U(1)_\text{A}$ operator with a fractional quantum Hall state to construct a \textit{conserved} and \textit{gauge-invariant} topological symmetry operator, denoted by ${\cal D}_{p\over N}(M)$, which can be supported on any closed, oriented three-manifold  $M$. 
The 2+1d fractional quantum Hall state is coupled to the bulk electromagnetic gauge field and only lives on the support of the operator. Therefore it does not change the bulk physics of QED. 

The price we pay, however, is that the topological operator ${\cal D}_{p\over N}$ does not obey a group multiplication law and does \textit{not} have an inverse operator ${\cal D}^{-1}_{p\over N}$ such that ${\cal D}_{p\over N} \times {\cal D}_{p\over N}^{-1}=1$.   
In particular, it is not a unitary operator. 
Can we still think of it as a global symmetry?

This question echoes with the recent developments of a novel kind of generalized global symmetries.  
(See \cite{McGreevy:2022oyu,Cordova:2022ruw} for reviews.)
While every ordinary global symmetry is associated with a topological symmetry operator (such as \eqref{introU}), the converse is not true. 
Building on earlier work of \cite{Verlinde:1988sn,Petkova:2000ip,Fuchs:2002cm,Frohlich:2004ef,Frohlich:2006ch,Frohlich:2009gb} in two spacetime dimensions, it has been advocated in \cite{Bhardwaj:2017xup,Chang:2018iay} that these more general topological operators should be viewed as generalized global symmetries. 
Since they do not have an inverse, they are commonly referred to as \textit{non-invertible symmetries}.
Some of the arguments for this interpretation  include: (1) Just like ordinary symmetries, some non-invertible symmetries can be gauged \cite{Frohlich:2009gb,Carqueville:2012dk,Brunner:2013xna,Bhardwaj:2017xup}. (2) When there is an obstruction  to gauging, the generalized 't Hooft anomaly matching condition leads to surprising constraints on renormalization group flows \cite{Chang:2018iay,Thorngren:2019iar,Komargodski:2020mxz,Thorngren:2021yso}. 
(3) In the context of quantum gravity, the no global symmetry conjecture is argued to be generalized to the absence of invertible and non-invertible symmetries \cite{Rudelius:2020orz,Heidenreich:2021xpr,McNamara:2021cuo,Arias-Tamargo:2022nlf}.  (4) Some non-invertible symmetries arise from invertible fermionic symmetries under bosonization \cite{Thorngren:2018bhj,Ji:2019ugf,Lin:2019hks,Burbano:2021loy,Kaidi:2021xfk}. 
The rapid developments of non-invertible global symmetries have led to many new constraints on strongly-coupled quantum systems in diverse spacetime dimensions, uncovering the more general algebraic and categorical structure of generalized global symmetries.

From this modern point of view, the new topological operators ${\cal D}_{p\over N}$ in QED are some of the first examples of non-invertible symmetries realized in Nature. 
They give an  invariant characterization of the ABJ anomaly in terms of the \textit{existence} of a generalized global symmetry, rather than the \textit{absence} thereof. 

In the past year, non-invertible symmetries have been constructed in many familiar  continuum and lattice gauge theories in higher than two spacetime dimensions \cite{Koide:2021zxj,Choi:2021kmx,Kaidi:2021xfk,Roumpedakis:2022aik,Bhardwaj:2022yxj,Hayashi:2022fkw,Choi:2022zal,Kaidi:2022uux}. 
These non-invertible symmetries are realized by gauging a higher-form global symmetry in either half of the spacetime \cite{Choi:2021kmx,Kaidi:2021xfk,Hayashi:2022fkw,Choi:2022zal}, or on  a higher-codimensional submanifold \cite{Roumpedakis:2022aik}.  
The operators ${\cal D}_{p\over N}$ in QED in this paper are realized from generalizations of such gauging constructions.

We further extend our analysis to QCD of the first generation in the massless limit.  
Below the electroweak scale, QCD has a $U(1)$  symmetry suffering from the ABJ anomaly with the electromagnetic gauge symmetry, which we now interpret as  an infinite, discrete, non-invertible symmetry.
Just like the ordinary global symmetry and anomaly, this non-invertible symmetry should be matched under renormalization group flows.  
We demonstrate that the coupling $\pi^0 F\wedge F$ in the IR pion Lagrangian is necessary to match the non-invertible symmetry in the QCD Lagrangian. 
Therefore, we have reinterpreted the conventional argument for the neutral pion decay using the ABJ anomaly as a matching condition for a non-invertible global symmetry.

\section{QED}

\subsection{ABJ Anomaly and the Fractional Quantum Hall State}\label{sec:ABJ}

Consider  QED of a unit charge, massless  Dirac fermion $\Psi$. The Euclidean Lagrangian is
\ie
{\cal L}_\text{QED}[\Psi,\bar\Psi , A] = \frac {1}{4e^2} F_{\mu\nu}F^{\mu\nu} + i \bar\Psi (\partial_\mu-i A_\mu) \gamma^\mu \Psi \,,
\fe 
where $A_\mu$ is the dynamical (compact) $U(1)$ one-form gauge field.  
We normalize the gauge field such that the flux $\oint_\Sigma F \in 2\pi \mathbb{Z}$ is properly quantized for any closed two-manifold $\Sigma$. 

Classically, there is a $U(1)_\text{A}$ axial global symmetry that acts on the fermion as
\ie\label{chiralrotation}
\Psi \to e^{i  \alpha \gamma_5/2} \Psi\,.
\fe
The normalization in the exponent is chosen in such a way that the periodicity of the axial rotation angle $\alpha$ is $2\pi$, i.e., 
 $\alpha\sim \alpha+2\pi$. 
 This is because the $\alpha=2\pi$ axial rotation acts on the  fermion as $\Psi \to e^{ i\pi \gamma_5} \Psi = (-1)\Psi$, which is part of the $U(1)$ gauge symmetry and is therefore a trivial transformation.
 
Quantum mechanically, the $U(1)_\text{A}$ axial symmetry is explicitly broken by the ABJ anomaly \cite{PhysRev.177.2426,Bell:1969ts}.
Let the axial current be
\begin{equation}
    j^\text{A}_{\mu} =  \bar{\Psi} \gamma_5  \gamma_\mu \Psi\,.
\end{equation}
Its conservation equation is violated by the dynamical gauge field, i.e., $\partial^\mu j_\mu^\text{A} = {1\over 16\pi^2} \epsilon^{\mu\nu\rho\sigma}F_{\mu\nu}F_{\rho\sigma}$. 
In terms of differential forms, we have
\begin{equation} \label{Eq:ABJ_anomaly}
d \star j^\text{A}  = \frac{1}{4\pi^2} F \wedge F \,.
\end{equation}

One can still attempt to define a $U(1)_\text{A}$  operator 
\begin{equation}
    U_\alpha (M) = \exp \left(
         \frac{i\alpha}{2} \oint_M \star j^\text{A} 
    \right) \,,
\end{equation}
where $M$ is a closed, oriented three-dimensional submanifold in spacetime on which the operator is supported. 
When $M$ is the whole space at a fixed time, the ABJ anomaly \eqref{Eq:ABJ_anomaly}  implies that this naive $U(1)_\text{A}$ symmetry operator  is not conserved under time evolution. 
In a relativistic QFT such as QED, \eqref{Eq:ABJ_anomaly} further implies that $U_\alpha(M)$ is generally not  topological.

Consider instead the combination 
\begin{equation}
    \star \hat{j}^\text{A}\equiv \star j^\text{A} - \frac{1}{4\pi^2} A\wedge dA 
\end{equation}
as a new current, which now formally satisfies the conservation equation $d \star \hat{j}^\text{A} = d\star j^\text{A} - \frac{1}{4\pi^2} F\wedge F = 0$.  
In components, we have $\hat j^\text{A} _\mu\equiv j^\text{A}_\mu -  {1\over 4\pi^2} \epsilon_{\mu\nu\rho\sigma} A^\nu \partial^\rho A^\sigma$. 
However, this new current is not gauge-invariant. 
It appears that there is no way to restore the $U(1)_\text{A}$ symmetry.

Nonetheless, let us naively proceed and  define a gauge non-invariant symmetry operator as
\begin{equation} \label{Eq:naive_symm_op}
    \hat{U}_\alpha(M) = \exp \left[
        \frac{i \alpha}{2} \oint_M \left( \star j^\text{A} - \frac{1}{4\pi^2} A\wedge dA  \right)
    \right] \,.
\end{equation}
Since the Chern-Simons level of $A\wedge dA$ in  \eqref{Eq:naive_symm_op} is not quantized, the exponent is not invariant under large gauge transformations when $M$ is a general compact three-manifold in spacetime.  
The operator $\hat{U}_\alpha (M)$ would have been topological, but it is  generally not well-defined  since it is not gauge-invariant.

Interestingly, there is a simple modification of \eqref{Eq:naive_symm_op}  for rational angles $\alpha \in 2\pi \mathbb{Q}$.  
Let us start with the  simplest case where $\alpha = 2\pi/N$ for some  integer $N$. 
In this case, the gauge non-invariant term in $\hat U_{2\pi\over N}(M)$ is
\ie\label{FQH1}
 - {i\over 4\pi N } \oint_M A\wedge dA\,.
\fe
Roughly speaking, this is the action for the  fractional quantum Hall state in 2+1d   at filling fraction $\nu=1/N$. 
(In that context, $A$ is regarded as a classical $U(1)$ background gauge field, whereas in the current context $A$ is a dynamical gauge field in the bulk.) 
However, this action on $M$ is not well-defined due to the fractional Chern-Simons level. 
Fortunately, there is a well-known solution to this inaccuracy in the condensed matter physics literature. (See, for example, \cite{Tong:2016kpv} for a review.) 
Instead of \eqref{FQH1}, the precise gauge-invariant action for the fractional quantum Hall state is
\ie\label{FQH2}
i\oint_M\left( \frac{N}{4\pi} a\wedge da + \frac{1}{2\pi} a\wedge dA\right)\,,
\fe
where $a$ is a dynamical $U(1)$ gauge field on $M$. 
It is a $U(1)_N$ Chern-Simons theory of $a$ coupled to   $A$. 
Integrating out $a$ naively gives us $a = -A/N$, which upon substitution returns  \eqref{FQH1}. 
However, this is not a rigorous equation since $-A/N$ is not a properly quantized $U(1)$ gauge field.  
It is therefore more precise to take \eqref{FQH2} as the action for the fractional quantum Hall state. 

Motivated by this discussion of the fractional quantum Hall state, we define a new operator ${\cal D}_{1\over N}$ in QED by replacing \eqref{FQH1} in $\hat U_{2\pi\over N}(M)$ with \eqref{FQH2}:
\begin{equation} \label{Eq:noninv_defect}
    \mathcal{D}_{1\over N} (M) = \exp \left[
            i \oint_M \left( \frac{2\pi}{2N} \star j^\text{A} + \frac{N}{4\pi} a\wedge da + \frac{1}{2\pi} a\wedge dA  \right)
        \right] \,,
\end{equation}
where $a$ is a dynamical one-form gauge field that only lives on the three-manifold $M$.\footnote{Here  and throughout we omit the path integral over $a$ in the expression for ${\cal D}_{1\over N}(M)$. }  
This new operator  can be viewed as dressing the naive axial symmetry operator $U_{2\pi \over N}(M)$ by  a fractional quantum Hall state on $M$ coupled to the bulk dynamical gauge field $A$.\footnote{Our operator is reminiscent of  the 2+1d sheet for the $\eta'$ particle in QCD in \cite{Komargodski:2018odf}. } 
We emphasize  that since $a$ only lives on the support of the operator $\mathcal{D}_\frac{1}{N}$, it can be viewed as an auxiliary field which does not change the physics of the bulk QED; in particular, there is no additional asymptotic state introduced by $a$.

The operator ${\cal D}_{1\over N}$ is distinguished from the previous trials in that it satisfies all the properties below:
\begin{itemize}
\item It acts as an axial rotation  on fermions with $\alpha=2\pi /N$ in \eqref{chiralrotation}.
\item It is gauge-invariant since the Chern-Simons levels are properly quantized.
\item It is topological, and in particular conserved under time evolution.
\end{itemize}
We will give a rigorous proof on the topological nature of (\ref{Eq:noninv_defect}) in Section \ref{sec:gauging}. 
For now, we can understand it heuristically from the relation between \eqref{FQH1} and \eqref{FQH2}, and the anomalous conservation equation \eqref{Eq:ABJ_anomaly}.

Since ${\cal D}_{1\over N}$ is a topological operator, it should be viewed as a generalized global symmetry in the spirit of \cite{Gaiotto:2014kfa,Bhardwaj:2017xup,Chang:2018iay}. 
Interestingly, it is not a usual group-like symmetry. 
That is, this operator doesn't follow the group multiplication law under parallel fusions. 
In particular, ${\cal D}_{1\over N}$ is not a unitary operator and it does  not have an inverse operator $({\cal D}_{1\over N})^{-1}$ such that ${\cal D}_{1\over N} \times ({\cal D}_{1\over N})^{-1}=({\cal D}_{1\over N})^{-1} \times {\cal D}_{1\over N}=1$.  
For this reason, ${\cal D}_{1\over N}$ is a   \emph{non-invertible symmetry}.
We will demonstrate some of these non-invertible fusion algebras  in Section \ref{sec:algebra}.

 \begin{table}
    \centering
    \begin{tabular}{c|c|c|c}
        & $U_\alpha(M)$ & $\hat{U}_\alpha(M)$ & $\mathcal{D}_{p\over N}(M)$  \\
        \hline
        Conserved (Topological) & \xmark & \cmark & \cmark \\
        \hline
        Gauge-invariant & \cmark & \xmark & \cmark  \\
        \hline
        Invertible & N/A & \cmark & \xmark  \\
        \hline
    \end{tabular}
    \caption{The non-invertible symmetry operator $\mathcal{D}_{p\over N}$ is both conserved (topological) and gauge-invariant, but it  does not obey a   group multiplication law under parallel fusion.  In contrast, the operator $\hat{U}_\alpha$ is not gauge-invariant, and  $U_\alpha$ is  not conserved due to the ABJ anomaly. (Since $U_\alpha$ is not topological, its fusion is subject to short-distance singularities and it is not meaningful to discuss its invertibility.)}
\end{table}

How do we generalize this construction to any rational angle $\alpha = 2\pi p/N$ with gcd$(p,N)=1$? 
In particular, how do we generalize \eqref{FQH2} for $p\neq 1$? 
A natural generalization of the $U(1)_N$ Chern-Simons theory   is the minimal $\mathbb{Z}_N$ TQFT ${\cal A}^{N,p}$ \cite{Hsin:2018vcg} (see also \cite{Moore:1988qv,Bonderson:2007ci,Barkeshli:2014cna}). 
The defining feature of ${\cal A}^{N,p}$ is that it is the minimal TQFT with a $\mathbb{Z}_N^{(1)}$ one-form global symmetry with its 't Hooft anomaly labeled by $p$. 
See  Appendix \ref{app:minimalTQFT} for a review of the minimal $\mathbb{Z}_N$ TQFT. 
When $p=1$, we have  ${\cal A}^{N,1}=U(1)_N$.

Let ${\cal A}^{N,p}[B]$ denote the Lagrangian of the $\mathbb{Z}_N$ minimal TQFT  coupled to   a $\mathbb{Z}_N$ background two-form gauge field $B$ for the $\mathbb{Z}_N^{(1)}$ one-form global symmetry. 
The natural  generalization of  the Lagrangian of \eqref{FQH2} is    ${\cal A}^{N,p}[dA/N]$, where we activate the two-form background gauge field by the  electromagnetic one-form gauge field $A$, properly normalized. 
With all these preparations, the new topological operator ${\cal D}_{p\over N}$ associated with the axial rotation $2\pi p/N$ is defined as\footnote{This non-invertible symmetry from the ABJ anomaly is  independently  constructed in  \cite{Cordova:2022ieu}. }
\begin{equation}\label{Eq:noninv_defect_kN}
    \mathcal{D}_{p\over N} (M) = \exp \left[
         \oint_M \left( \frac{2\pi i p}{2N} \star j^\text{A} + {\cal A}^{N,p}[dA/N]   \right)
    \right] \,.
\end{equation}
We will give a more detailed justification in Appendix \ref{app:minimalTQFT}.  
Since ${\cal A}^{N,p+N} ={\cal A}^{N,p}$, we have ${\cal D}_{p+N\over N}={\cal D}_{p\over N}$, and therefore the non-invertible symmetry is labeled by an element $p/N  \in \mathbb{Q}/\mathbb{Z}$.

We can replace ${\cal A}^{N,p}[B]$ in \eqref{Eq:noninv_defect_kN} by any 2+1d TQFT $\cal T[B]$ (e.g., $p$ copies of $U(1)_N$) with a $\mathbb{Z}_N^{(1)}$ one-form symmetry and anomaly  $p$. 
This defines another topological operator ${\cal D}_{\cal T}$. 
It was shown in \cite{Hsin:2018vcg} any such TQFT ${\cal T}[B]$ is factorized as ${\cal T}[B] = {\cal A}^{N,p}[B] \otimes {\cal T}'$, where ${\cal T}'$ is  a decoupled TQFT. 
It follows that ${\cal D}_{\cal T}$ is a composite operator of ${\cal D}_{p\over N}$ and a decoupled 2+1d TQFT ${\cal T}'$, i.e., ${\cal D}_{\cal T} =  {\cal D}_{p\over N}\times {\cal T}'$. 
In this sense, ${\cal D}_{p\over N}$ defined in \eqref{Eq:noninv_defect_kN} is the minimal topological operator.

Let us summarize our discussion so far. 
In massless QED, for every rational angle $\alpha = 2\pi p/N$, there is a gauge-invariant and conserved topological symmetry operator ${\cal D}_{p\over N}$ that acts on the fermions as axial rotations.  
However, there is no gauge-invariant Noether current or charge.  
Indeed, the exponents of \eqref{Eq:noninv_defect} and \eqref{Eq:noninv_defect_kN}, which would have been the conserved charges, are not gauge-invariant because of the Chern-Simons terms. 
Rather, their exponentiations are gauge-invariant symmetry operators. 
Therefore, the non-invertible symmetries from ${\cal D}_{p\over N}$ are discrete, rather than continuous. 
We conclude that the continuous, invertible axial $U(1)_\text{A}$ symmetry is broken by the ABJ anomaly to a discrete, non-invertible symmetry.

Finally, we comment on the non-invertible symmetry ${\cal D}_{p\over N}$ and  $\hat U_\alpha$ in \eqref{Eq:naive_symm_op} on non-compact space such as $\mathbb{R}^3$. 
(See, for example,  \cite{Harlow:2018tng} for  recent discussions.)
In this case, the operator $\hat U_\alpha(\mathbb{R}^3)$ is actually gauge-invariant   because there is no non-trivial gauge transformation on $\mathbb{R}^3$ or $S^3$  since $\pi_3(U(1))=0$. 
In fact, on $\mathbb{R}^3$, we can integrate out $a$ in \eqref{Eq:noninv_defect_kN} and equate ${\cal D}_{p\over N}  (\mathbb{R}^3) = \hat U_{2\pi p\over N}(\mathbb{R}^3)$. 
However, $\hat U_\alpha(M)$ is not gauge-invariant  on  a  more general compact three-manifold $M$. 
In contrast, our non-invertible symmetry ${\cal D}_{p\over N}(M)$ \eqref{Eq:noninv_defect_kN} is gauge-invariant and conserved (topological) for any compact three-manifold $M$, but it is only defined for rational angles $\alpha=2\pi p/N$.
 
\subsection{Non-invertible Fusion Algebra over TQFT Coefficients}\label{sec:algebra}

Having identified an infinite number of conserved symmetry  operators ${\cal D}_{p\over N}$, we now demonstrate some of their non-invertible fusion rules under parallel fusion, leaving the determination of the full fusion algebra to future work. The general structure of the fusion of two such operators ${\cal D},{\cal D}'$ takes the following form:
\ie
{\cal D} (M) \times {\cal D}'(M)  =  {\cal T}(M) {\cal D}''(M)\,,
\fe
where ${\cal D}''$ is another conserved symmetry operator, and ${\cal T}(M)$ is the fusion ``coefficient." 
Surprisingly, the fusion ``coefficient" is generally not a number, but a 2+1d TQFT $\cal T$. 
(More precisely, the fusion coefficient is the partition function of the TQFT $\cal T$ evaluated on the three-manifold $M$.) 
Similar fusion algebras over TQFT coefficients have recently been explored in \cite{Roumpedakis:2022aik,Choi:2022zal}.

For an invertible global symmetry, its symmetry operator $U_g$ is unitary, i.e., $U_g(M)\times (U_g)^\dagger(M) = U_g(M)\times U_{g^{-1}}(M)= 1$.\footnote{More generally, when $M$ is not a spatial slice at a fixed time, the $\dagger$ should be replaced by the orientation reversal of an operator/defect  defined in \cite{Roumpedakis:2022aik}.  The orientation reversal $\overline{\cal D}$ of an operator/defect ${\cal D}$ is defined by $\overline{\cal D}({\overline M})  = {\cal D}(M)$, where $\overline M$ is the orientation reversal of the three-manifold $M$.}
In contrast, our non-invertible topological symmetry operators are not unitary. 
The $\dagger$ of the operator ${\cal D}_{1\over N}$ is
\ie
{\cal D}^\dagger_{1\over N}(M)  = \exp \left[
           - i \oint_M \left( \frac{2\pi}{2N} \star j^\text{A} + \frac{N}{4\pi}\bar a\wedge d \bar a + \frac{1}{2\pi}\bar a\wedge dA  \right)
        \right] \,,
\fe
where we have used a different symbol $\bar a$ to denote the dynamical gauge field living on ${\cal D}^\dagger_{1\over N}(M)$ to distinguish from the one in \eqref{Eq:noninv_defect}. 
Note that ${\cal D}_{1\over N}^\dagger (M)={\cal  D}_{-{1\over N}}(M)$. 
The fusion of ${\cal D}_{1\over N}$ and ${\cal D}_{1\over N}^\dagger$ is then 
\ie\label{DDbar}
{\cal D}_{1\over N}(M)\times {\cal D}_{1\over N}^\dagger (M) 
= \exp \left[
            i \oint_M \left(  \frac{N}{4\pi} a\wedge da  -  \frac{N}{4\pi} \bar a\wedge d\bar a + \frac{1}{2\pi} (a-\bar a)\wedge dA  \right)
        \right] \equiv {\cal C} \neq1\,.
\fe
We see that while the naive axial rotation component  is unitary, the additional fractional quantum Hall state, which is necessary for the conservation, is not unitary.  
This is  a generalization of the phenomenon observed in \cite{Kaidi:2021xfk} for discrete ABJ-type anomalies. 

The operator on the RHS of \eqref{DDbar} is  an example of the \textit{condensation operator} $\cal C$, which has recently drawn a lot of attention in the literature \cite{Kong:2014qka,Else:2017yqj,Gaiotto:2019xmp,Kong:2020cie,Johnson-Freyd:2020twl,Roumpedakis:2022aik,Choi:2022zal} (see also \cite{Choi:2021kmx,Kaidi:2021xfk}).  
More specifically, it is the magnetic condensation operator from the higher gauging \cite{Roumpedakis:2022aik} of a discrete subgroup of the magnetic one-form global symmetry. 
We will have a more detailed discussion in Appendix \ref{app:Maxwell}.

Let us compute another  fusion product:
\ie
{\cal D}_{1\over N}(M)\times {\cal D}_{1\over N} (M)=\exp\left[ i \oint_M \left(
{4\pi \over 2N} \star j^\text{A}+
 \frac{N}{4\pi} a_1\wedge da_1+\frac{N}{4\pi} a_2\wedge da_2+\frac{1}{2\pi}(a_1+a_2)\wedge dA  \right)\right] ~.
\fe
If $N$ is odd, we can decompose the $U(1)_N\times U(1)_N$ Chern-Simons theory of $a_1, a_2$ on $M$ into two minimal TQFTs, $\mathcal{A}^{N,2}\times\mathcal{A}^{N,2}$, generated by the Wilson line $\exp(i\oint (a_1+a_2))$ and $\exp(i\oint (a_1-a_2))$, respectively. 
(See Appendix \ref{app:minimalTQFT} for more details of this decomposition.) 
Only the first minimal theory is coupled to $A$, while the other copy is decoupled. Hence,
\ie
\mathcal{D}_{1\over N} \times\mathcal{D}_{1\over N}=\mathcal{A}^{N,2} \,\mathcal{D}_{2\over N}~,~~~\text{odd}~N\,.
\fe
A similar fusion rule was recently reported in \cite{Choi:2022zal} in the context of ${\cal N}=1$ super Yang-Mills theory.

\subsection{Gauging the Magnetic One-Form Symmetry}\label{sec:gauging}

In this subsection we give   an alternative construction of the non-invertible symmetry ${\cal D}_{p\over N}$ by gauging a one-form global symmetry. 
This construction gives  a rigorous proof of the conservation, or more generally the topological nature, of $\mathcal{D}_{p\over N}$ defined in \eqref{Eq:noninv_defect_kN}. 
In this subsection we will view ${\cal D}_{p\over N}$ as  a defect that is supported on the $x=0$ three-manifold in spacetime. 

QED has a magnetic one-form global symmetry $U(1)^{(1)}$, whose conserved Noether current is  a two-form, $j^\text{m} \equiv {1\over 2\pi} \star F$ \cite{Gaiotto:2014kfa}.\footnote{In \cite{Gaiotto:2014kfa}, the Noether current of a $q$-form symmetry is defined as a $(d-q-1)$-form current, where $d$ is the spacetime dimensions. The Noether currents in this paper  are the Hodge duals of those in \cite{Gaiotto:2014kfa}.} 
The conservation equation simply follows from the Bianchi identity, i.e., $d\star j^\text{m} = {1\over 2\pi} d F= 0$. 
The charged objects under this magnetic one-form global symmetry are the 't Hooft lines. 
We will assume the absence of  dynamical monopoles, otherwise the magnetic one-form symmetry is broken and our proof below will not hold.  
See  \cite{Cordova:2022ieu} for discussions on the breaking of the non-invertible symmetry by dynamical monopoles.

The background gauge field for the magnetic one-form global symmetry $U(1)^{(1)}$ is a two-form gauge field $B$.  
It is coupled to the QED Lagrangian by $ i B\wedge \star j^\text{m} = {i\over 2\pi } B\wedge F$. 
In addition, one can add  local counterterms that depend only on the background gauge field $B$.

Below we will gauge the $\mathbb{Z}_N^{(1)}$ subgroup of the $U(1)^{(1)}$ symmetry. 
We first promote the two-form background gauge field $B$ to be a dynamical gauge field and denote the latter as $b$. 
Next, we introduce a dynamical $U(1)$ one-form gauge field $c$ that couples to $b$ as
\begin{equation} \label{Eq:bdc}
    \frac{iN}{2\pi} b \wedge dc \,.
\end{equation}
The equation of motion for the Lagrange multiplier field $c$ forces $b$ to be flat, i.e.,  $db = 0$ and restricts holonomy of $b$  to be $\mathbb{Z}_N$-valued, making $b$ a $\mathbb{Z}_N$ two-form gauge field. 
The remaining path integral over $b$ collapses to a finite sum which implements the gauging of the $\mathbb{Z}_N^{(1)}$ subgroup of the magnetic one-form symmetry. 
See \cite{Maldacena:2001ss,Banks:2010zn,Kapustin:2014gua} for detailed discussions of this presentation \eqref{Eq:bdc} of discrete gauge theory.

Finally, we can add an additional term that only depends on  the two-form gauge field $b$ when we gauge the $\mathbb{Z}_N^{(1)}$ magnetic one-form symmetry. 
This is the discrete analog of a Maxwell action, which is known as a discrete torsion in high energy physics or as a Symmetry Protected Topological (SPT) phase in condensed matter physics. 
For our purpose, we choose this term to be ${i Nk\over 4\pi} b\wedge b$, where $k$ is the multiplicative inverse of $p$ modulo $N$, i.e., $pk =1$ mod $N$.\footnote{Such an integer $k$ always exist since gcd$(p,N)=1$.} 
Putting everything together,  gauging the $\mathbb{Z}_N^{(1)}$ magnetic one-form symmetry of QED is described by the following Lagrangian\footnote{Here and below we slightly abuse the notation and treat the QED Lagrangian ${\cal L}_\text{QED}[\Psi,\bar\Psi,A]$ as a four-form rather than a scalar.}
\begin{equation}\label{discretegauging}
    \mathcal{L}_\text{QED}[\Psi,\bar{\Psi},A] + \frac{i}{2\pi} b \wedge F + \frac{iN}{2\pi} b \wedge dc
    + \frac{iNk}{4\pi} b\wedge b \,.
\end{equation}
In Appendix \ref{app:Maxwell}, we will discuss in more details the meaning of this discrete gauging and associate this gauging with an element, denoted as $CT^{(k)} S^{(k)} T^{(k)}$, of the modular group.  
As shown in that appendix, this discrete gauging shifts the $\theta$-angle for the $U(1)$ gauge field by 
\ie
\theta \mapsto \theta - 2\pi p/N\,.
\fe
In QED, this shift of the $\theta$-angle can be undone by an axial rotation \eqref{chiralrotation} of the fermions with $\alpha = 2\pi p/N$. 
We thus conclude that the massless QED is invariant under the discrete gauging \eqref{discretegauging}.

Whenever a QFT is invariant under gauging a discrete global symmetry, we can define a codimension-one topological operator/defect by gauging only in half of the spacetime \cite{Kaidi:2021gbs,Choi:2021kmx,Choi:2022zal}. 
More specifically, we proceed as follows:
\begin{itemize}
\item First, we apply  the discrete gauging \eqref{discretegauging} in the $x\geq 0$ region.
\item Next, we add to the total Lagrangian the following: 
\ie\label{step2}
{2\pi i p\over 2N} \oint_{x=0} \star j^\text{A} +{ip\over 4\pi N} \int_{x\ge0} F\wedge F \,.
\fe
Using the anomalous conservation equation  \eqref{Eq:ABJ_anomaly}, the above combination is trivial, so it is justified to add it without any cost. 
These terms can also be understood from   a change of variables on the  fermions by a spacetime-dependent axial rotation $\Psi (x)\to \exp( i \gamma_5 \alpha(x)  /2) \Psi(x)$. 
This change of variables shifts the action by $S \to S - \frac i2 \int d^4x j^\text{A}_\mu(x)\partial^\mu \alpha(x)$ and the measure by $[D\Psi][D\bar\Psi] \to [D\Psi] [D\bar\Psi] \exp \left[ {i\over 8\pi^2} \int  \alpha(x)F\wedge F\right]$ \cite{PhysRevD.29.285}. 
If we choose $\alpha(x) = {2\pi p \over N} \Theta(x)$, where $\Theta(x)=1$ if $x\ge0$ and $\Theta(x)=0$ if $x<0$, then it gives the two terms in \eqref{step2}.
\end{itemize}

Composing the discrete gauging and the axial rotation in the $x\ge 0$ region, the total Lagrangian becomes
\ie\label{Eq:TST_interface}
&    \int_{x<0} \mathcal{L}_\text{QED}[\Psi,\bar{\Psi},A] 
    +{2\pi i p\over 2N} \oint_{x=0} \star j^\text{A}\\
  &  + \int_{x\geq 0} \left(
    \mathcal{L}_\text{QED}[\Psi,\bar{\Psi},A] + \frac{ip}{4\pi N} F \wedge F + \frac{i}{2\pi} b \wedge F + \frac{iN}{2\pi} b \wedge dc + \frac{iNk}{4\pi} b \wedge b \right) \,.
\fe
Importantly, the dynamical gauge fields $b,c$ are only defined in the $x\ge 0$ region, and we impose the Dirichlet boundary condition $b|_{x=0} = 0$ at $x=0$.
This defect is manifestly topological since the Dirichlet boundary condition for the two-form gauge field $b$ is topological, which in turn follows from  the flatness condition $db =0$. (See \cite{Choi:2021kmx} for a more detailed explanation on this topological boundary condition.)

In Appendix \ref{app:minimalTQFT}, we show that the path integral over $b,c$ in the $x\ge0$ region gives 
\begin{equation} \label{Eq:minimal_bulk}
    \int_{x\geq 0} \left(
    {ip\over 4\pi N} F\wedge F+
    \frac{i}{2\pi} b \wedge F + \frac{iN}{2\pi} b \wedge dc + \frac{iNk}{4\pi} b \wedge b \right)
    = \oint_{x=0} \mathcal{A}^{N,p}[dA/N] \,.
\end{equation}
Therefore, the bulk Lagrangians in the $x<0$ and $x>0$ regions are both the original QED Lagrangian, but the discrete gauging  leaves behind a defect at $x=0$:
\ie
\oint_{x=0}\left(  {2\pi i p\over 2N} \star j^\text{A}  + {\cal A}^{N,p}[dA/N] \right)
\fe
which is exactly the defect ${\cal D}_{p\over N}$ in \eqref{Eq:noninv_defect_kN}. 
This alternative construction via  discrete gauging proves the topological nature of ${\cal D}_{p\over N}$ in QED.

Finally, we comment that the above construction of the non-invertible symmetry applies to any abstract 3+1d QFT with the following properties:
\begin{itemize}
\item It has a $U(1)^{(1)}$ one-form global symmetry with a conserved two-form current $j^\text{m}$:
\ie
d\star j^\text{m}=0\,.
\fe
\item It has a spin-one operator $j^\text{A}$  obeying the anomalous conservation equation:
\ie
d\star j^\text{A} = \star j^\text{m}\wedge \star j^\text{m}\,.
\fe
\end{itemize}
Then we can follow the identical steps to construct the non-invertible  symmetry 
\ie
{\cal D}_{p\over N}(M)= \exp\left[ \oint_M\left(  {2\pi i p\over 2N} \star j^\text{A}  + {\cal A}^{N,p}\left[{2\pi\over N}\star j^\text{m}\right] \right)\right]\,.
\fe 
In particular, we do not need to assume that the QFT has a Lagrangian description in terms of fermions and gauge fields. 
 
\subsection{Action on Operators and Selection Rules}

In Section \ref{sec:gauging}, we realize the operator ${\cal D}_{p\over N}$ as a composition of an axial rotation and gauging a $\mathbb{Z}_N^{(1)}$ subgroup of the magnetic one-form symmetry in half of the spacetime.  
We can determine the action of ${\cal D}_{p\over N}$ on the local and the line operators from this gauging construction.

Since the fermions are not affected by gauging the magnetic one-form symmetry, ${\cal D}_{p\over N}$ acts invertibly on them as an axial rotation with a rational angle $\alpha=2\pi p/N$. 
This is also clear from the first term in \eqref{Eq:noninv_defect_kN}.  
Similarly, the non-invertible symmetry acts trivially on the Wilson lines. 

In particular, a Dirac mass term $m\bar\Psi \Psi$ in the Lagrangian for the electron would violate the non-invertible global symmetry ${\cal D}_{p\over N}$. 
In this sense, we can say that the \textit{electron is naturally massless} in QED because of the non-invertible symmetry.  
In the literature, this naturalness is commonly attributed to the classical axial symmetry and the fact that there is no $U(1)$ instanton in flat spacetime \cite{tHooft:1979rat}. 
We have provided an alternative explanation using an exact non-invertible global symmetry in massless QED.

However, we will see that its action on the 't Hooft line  is more subtle. 
Let us denote the minimal 't Hooft line on a closed curve $\gamma$ as $H(\gamma)$.
The background gauge transformation of the $U(1)^{(1)}$ magnetic one-form symmetry acts on the minimal 't Hooft line $H(\gamma)$ and the background two-form gauge field $B$ as 
\ie
H(\gamma)\mapsto  H(\gamma) e^{i \oint_\gamma \Lambda}\,,~~~~B\mapsto B+d\Lambda
\fe
 where $\Lambda$ is the one-form gauge parameter.   
When we  gauge the $\mathbb{Z}_N^{(1)}$ magnetic one-form symmetry as in \eqref{discretegauging}, the minimal 't Hooft line is no longer gauge-invariant. 
Rather, when $\gamma$ is a contractible loop, the gauge-invariant operator is 
\ie
 H(\gamma) \exp\left(- i \int_\Sigma b\right) \,,
 \fe
where $\Sigma$ is a two-dimensional disk such that $\partial\Sigma = \gamma$.  
Next, we integrate out $c$ in \eqref{discretegauging}, which constrains $b$ to be a $\mathbb{Z}_N$ gauge field. 
 Then the equation of motion of $b$ gives    (see Appendix \ref{app:zerogauging} for more details)
 \ie\label{disk}
 H (\gamma)\exp\left( i {p\over N}\int_\Sigma F \right) \,.
 \fe

\begin{figure}[t]
    \centering
    \begin{subfigure}[T]{.45\textwidth}
        \centering
        \includegraphics[width=0.9\linewidth]{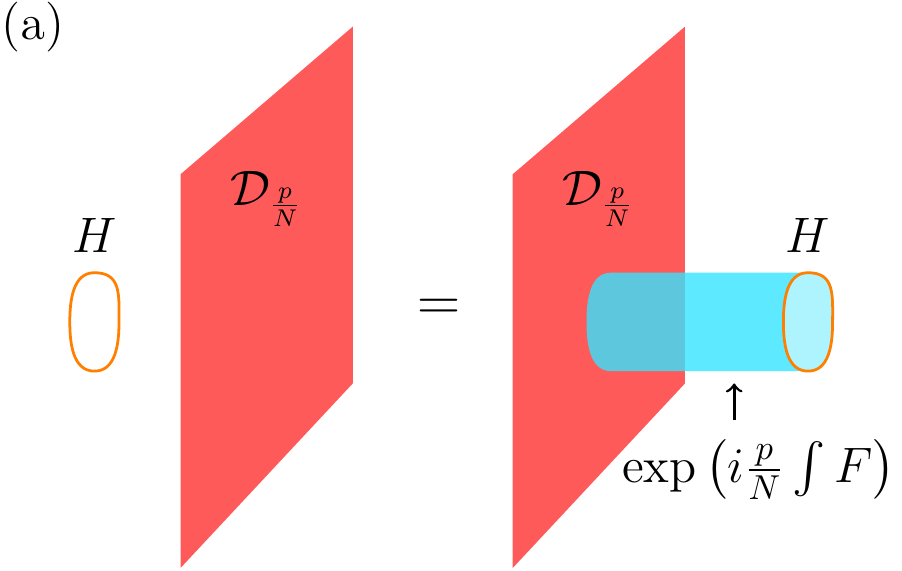}
    \end{subfigure}
    \hspace{0.5cm}
    \begin{subfigure}[T]{.45\textwidth}
        \centering
        \includegraphics[width=0.9\linewidth]{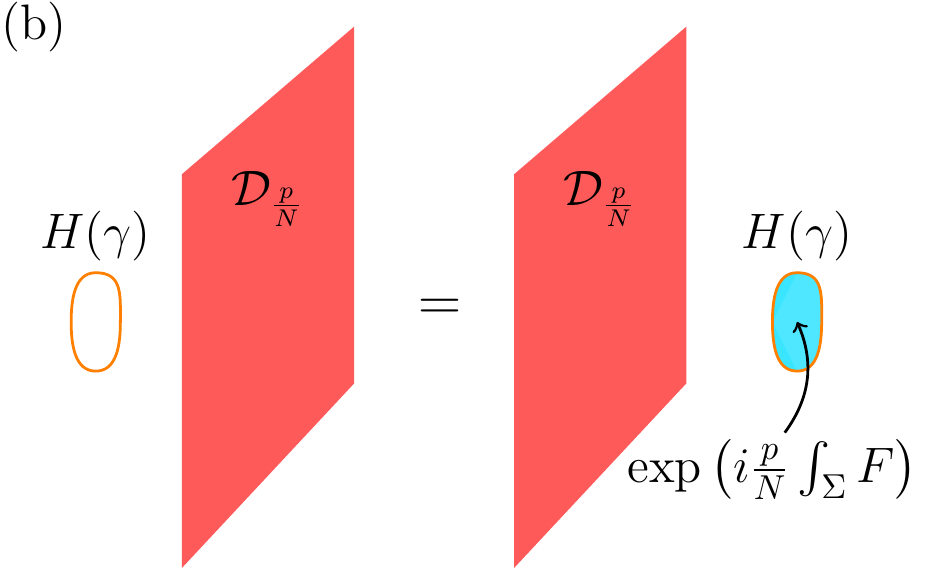}
    \end{subfigure}
    \caption{Action of the three-dimensional non-invertible symmetry operator $\mathcal{D}_{\frac{p}{N}}$ on the 't Hooft line $H$.
    (a) As one sweeps the operator $\mathcal{D}_{\frac{p}{N}}$ past the 't Hooft line $H$, the latter is attached to the topological surface operator $\exp \left( i \frac{p}{N} \int F \right)$ which is stretched between $\mathcal{D}_{\frac{p}{N}}$ and $H$.
    The whole configuration is gauge-invariant.
    (b) When the 't Hooft line is supported on a contractible loop $\gamma$ such that $\gamma= \partial \Sigma$ for a two-dimensional disk $\Sigma$, the surface operator $\exp \left( i \frac{p}{N} \int F \right)$ can be deformed to be supported on $\Sigma$.
    The operator $\exp \left( i \frac{p}{N} \int_\Sigma F \right)$ can be thought of as a fractional Wilson line on $\gamma$.}
    \label{Fig:4d}
\end{figure}

From this, we can deduce the action of the non-invertible symmetry operator $\mathcal{D}_{\frac{p}{N}}$ on the 't Hooft line $H(\gamma)$.
As we sweep the operator ${\cal D}_{p\over N}$ past the 't Hooft line, the latter is attached to the topological surface operator $\exp\left( i {p\over N}\int F \right)$ which is stretched between the symmetry operator $\mathcal{D}_{\frac{p}{N}}$ and the 't Hooft line $H(\gamma)$.
This configuration is gauge-invariant.
Moreover, if $\gamma$ is a contractible loop, we can topologically deform the surface operator $\exp\left( i {p\over N}\int F \right)$ to be supported on a disk $\Sigma$ which bounds $\gamma = \partial \Sigma$.
The latter can be thought of as an improperly quantized Wilson line with a fractional electric charge $p/N$.  
See Figure \ref{Fig:4d}. 
Similar transformations on the 't Hooft lines have been discussed, for example, in \cite{Harlow:2018tng}. 

This action of ${\cal D}_{p\over N}$ on the 't Hooft line can be understood from the Witten effect  \cite{Witten:1979ey} as follows.
In Appendix \ref{app:zerogauging}, we show that the $CT^{(k)} S^{(k)} T^{(k)}$ discrete gauging \eqref{discretegauging} shifts the $\theta$-angle as $\theta \mapsto \theta- 2\pi p/N$. 
According to the Witten effect, the 't Hooft line acquires a fractional electric charge $p/N$ and becomes \eqref{disk}.

\begin{figure}[t]
    \centering
    \includegraphics[width=0.3\textwidth]{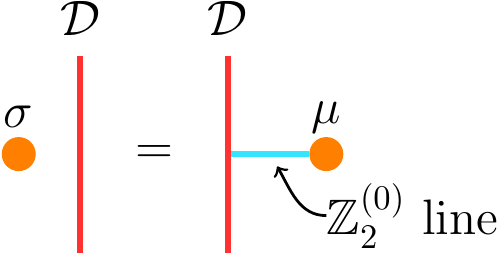}
    \caption{Action of the non-invertible Kramers-Wannier duality line $\mathcal{D}$ on the spin operator $\sigma$ in the 1+1d Ising CFT.
    When $\mathcal{D}$ sweeps past $\sigma$, the latter becomes the disorder operator $\mu$ attached to the topological $\mathbb{Z}_2^{(0)}$ line.}
    \label{Fig:2d}
\end{figure}

Let us draw an analogy with the non-invertible Kramers-Wannier duality line $\cal D$ in the 1+1d Ising CFT.
The line $\cal D$ successively implements the gauging of the $\mathbb{Z}_2^{(0)}$ zero-form symmetry of the Ising CFT and the Kramers-Wannier duality transformation only on half of the spacetime.
This is analogous to $\mathcal{D}_{\frac{p}{N}}$, which implements the $CT^{(k)} S^{(k)} T^{(k)}$ gauging and the axial rotation of fermions only on half of the spacetime.
As we sweep $\cal D$ past a local spin operator $\sigma$, the latter becomes a disorder operator $\mu$ that is attached to a $\mathbb{Z}_2^{(0)}$ line \cite{Frohlich:2004ef,Chang:2018iay}.
See Figure \ref{Fig:2d}.

Above we described the Euclidean configurations of 't Hooft lines and the non-invertible symmetries. What is the action of the operator ${\cal D}_{p\over N}$ on the Hilbert space? 
Consider specifically the state $|H(\gamma)\rangle$ obtained from  wrapping the 't Hooft line around a non-trivial one-cycle $\gamma$ in space. 
Since there is no way to fill in the one-cycle $\gamma$, the operator ${\cal D}_{p\over N}$ will annihilate this state in the Hilbert space, i.e., ${\cal D}_{p\over N}|H(\gamma)\rangle=0$. 
In this sense, the operator ${\cal D}_{p\over N}$ is non-invertible when acting on states created by the 't Hooft lines. 
Again, this is similar to the action of the Kramers-Wannier duality line $\cal D$ in the Ising CFT on the state $|\sigma\rangle$ corresponding to the spin operator, i.e., ${\cal D}|\sigma\rangle=0$ \cite{Petkova:2000ip,Chang:2018iay}.

Since the operators ${\cal D}_{p\over N}$ act invertibly on the fermions  as rational axial rotations \eqref{chiralrotation}, they lead to the same selection rule on flat space amplitudes as what the naive $U(1)_\text{A}$ symmetry would. 
The naive $U(1)_\text{A}$ selection rule also follows from the operator $\hat U_\alpha(\mathbb{R}^3)$ in \eqref{Eq:naive_symm_op}, which is gauge-invariant because there is no $U(1)$ instanton in flat spacetime. 
More specifically, since chirality is tied to the helicity for massless fermions, this selection rule requires that the sum of the electron and positron helicities have to be conserved. 
This is the familiar helicity conservation for the electrons and the positrons in massless QED.\footnote{In terms of the spinor helicity variables for the electrons and positrons, the non-invertible symmetry ${\cal D}_{p\over N}$ acts  as $|p] \to e^{-{2\pi i p\over N} }|p]\,,~~~|p\rangle \to e^{2\pi i p\over N} |p\rangle\,,~~~
[p| \to e^{-{2\pi i p\over N} } [p|\,,~~~\langle p| \to e^{{2\pi i p\over N} }\langle p|.$}  
For example, in the positron-electron annihilation $e^+e^- \to \gamma\gamma$, the incoming positron and electron have to carry opposite helicities. 
As another example, the total helicity is conserved in the Bhabha scattering $e^+ e^- \to e^+e^-$.

Importantly, the helicity conservation only applies to the electrons and positrons, but not to the photons. 
In fact, the sum of the photon helicities is generally not conserved in massless QED. 
For example,  the total photon helicity is violated at one-loop for the electron-positron annihilation into to $n$ positive helicity photons with $n\ge 4$ \cite{Mahlon:1993fe}.

\section{QCD and the Pion Decay}

Let us take the UV theory to be the QCD Lagrangian of the massless up and down quarks at an energy scale way above the pion scale, but below the electroweak scale so the  $SU(2)\times U(1)$ gauge symmetry has been Higgsed to the electromagnetic $U(1)_\text{EM}$ gauge symmetry. 
Let $u,d$ be the Dirac fermions for the up and down quarks, respectively. 
The $U(1)_\text{EM}$ charges of the $u$ and $d$ are $+2/3$ and $-1/3$, respectively. 
We will suppress the $SU(3)$ color indices. 
Classically, the QCD Lagrangian has a global symmetry\footnote{The subscript 3 is to distinguish this symmetry from the other  axial symmetry that acts as $ \left(\begin{array}{c}u  \\ d\end{array}\right) \to \exp\left( i  {\alpha\over 2} \gamma_5   \right) \left(\begin{array}{c}u \\d\end{array}\right)$, which suffers from an ABJ anomaly with the $SU(3)$ gauge symmetry. }
\ie
U(1)_{\text{A}3}:~ \left(\begin{array}{c}u\\d\end{array}\right)
\to
\exp\left( i  {\beta} \gamma_5 \sigma_3 \right) \left(\begin{array}{c}u \\d\end{array}\right)
=\left(\begin{array}{c} \exp\left( i  {\beta } \gamma_5\right) u\\ \exp\left( -i  {\beta } \gamma_5\right) d\end{array}\right)\,,
\fe
where $\beta\sim \beta+2\pi$. 
The axial current is conventionally normalized as
\ie
j^\text{A3}_{\mu}  = \frac 12 \overline{u} \gamma_5 \gamma_\mu u
-\frac 12 \overline d \gamma_5 \gamma_\mu d \,.
\fe
It suffers from the the ABJ anomaly with the electromagnetic $U(1)_\text{EM}$ gauge symmetry:
  \ie
 d\star j^\text{A3} = {1\over 8\pi^2} F\wedge F \,.
 \fe
 The naive, gauge non-invariant symmetry operator is  $\exp\left[
 2i \beta \oint_M \left( \star j^\text{A3} - {1\over 8\pi ^2} A\wedge dA\right)
 \right]$. 
 Note that for $\beta=\pi$,  the $-{i\over 4\pi}A\wedge dA$ term is actually properly quantized, and it is an invertible $\mathbb{Z}_2$ symmetry operator
 \ie
 \exp\left[
 \oint_M \left(2\pi i  \star j^\text{A3} - {i\over 4\pi } A\wedge dA\right)
 \right]
 \fe
  that acts as $u \to -u, d\to-d$.  
 In other words, the classical $U(1)_\text{A3}$ symmetry is broken to a $\mathbb{Z}_2$ global symmetry quantum mechanically.  
 
For more generic rational angle $\beta$, say, $\beta=\pi/N$, we can apply the same construction in Section \ref{sec:ABJ} to define a gauge-invariant and conserved topological operator
 \ie\label{DQCD}
{\cal D}_{1\over N}(M)= \exp\left[ i
 \oint_M \left( {2\pi   \over N}\star j^\text{A3} + {N\over 4\pi } a\wedge da + {1\over 2\pi }a\wedge dA\right)
 \right] \,.
 \fe
 The topological nature of this operator is proved using an identical argument as in Section \ref{sec:gauging} by gauging the $\mathbb{Z}_N^{(1)}$ subgroup of the $U(1)^{(1)}$ magnetic one-form symmetry.\footnote{Since the quarks have $+2/3$ and $-1/3$ charges, we should define $A '=A/3$, with $A'$ normalized by requiring that $\oint_\Sigma F'\in 2\pi \mathbb{Z}$ for every closed two-manifold $\Sigma$. This would imply that some of these non-invertible operators are not minimal and can be factorized into an invertible symmetry operator times another topological operator which acts trivially on all local operators. Nonetheless, they should still be matched under renormalization group flows.}

How are these infinitely many non-invertible symmetries in QCD captured by the low-energy pion Lagrangian? 
In the pion Lagrangian, the axial current becomes
 \ie
 j^\text{A3}_\mu =  - f_\pi \partial_\mu \pi^0 +\cdots\,,
 \fe
 which shifts the neutral pion by $\pi^0 \to \pi^0 -2\beta f_\pi$.  
 Here $f_\pi \sim 92.4$ MeV is the pion decay constant. 
 Since the neutral pion field is compact with periodicity $\pi^0 \sim \pi^0+2\pi f_\pi$, the $\beta=\pi$ transformation, which generates a $\mathbb{Z}_2$ global symmetry in QCD, now acts trivially in the IR pion Lagrangian.

The relevant terms in the pion Lagrangian in Euclidean signature are
\ie\label{IRL}
{\cal L}_\text{IR} = {1\over 2} d\pi^0 \wedge \star d\pi^0  +{1\over 2e^2} F\wedge \star F
+ i g \, \pi^0 F\wedge F + i A\wedge \star J^\text{EM} + \cdots\,,
\fe
where $J^\text{EM}_\mu =  \pi^-   \partial_\mu\pi^+ - \pi^+ \partial_\mu \pi^- +\cdots$ is the current that couples to the electromagnetic gauge field. 
Here $g$ is a coefficient we will fix by  the  non-invertible symmetry.

To proceed, we insert ${\cal D}_{1\over N}$ as  a  defect at $x=0$ in the IR pion effective theory. 
In other words, $M$ is chosen to be the three-manifold defined as $x=0$ in Euclidean spacetime. 
For this to be a consistent defect, we need to investigate the equations of motion. 
Because of the first term in \eqref{DQCD}, the pion field is discontinuous across the defect:
\ie\label{pidiscont}
\pi^0\Big|_{x=0^+} = \pi^0\Big |_{x=0^- } - {2\pi \over N}f_\pi\,. 
\fe
The equation of motion for the gauge field $a$ on $M$ gives
\ie\label{aeom}
Nda+F=0\,.
\fe
 On the other hand, the equation of motion for the bulk gauge field $A$ gives
 \ie
& \int_{x<0}   \delta A\wedge \left[
\left( - {1\over e^2}d \star F+i \star J^\text{EM}\right) - 2i g d (\pi^0 F)
 \right]
-{i\over 2\pi} \int_{x=0} \delta A\wedge da\\
&+ \int_{x>0}   \delta A\wedge \left[
\left(- {1\over e^2}d \star F+i \star J^\text{EM}\right) - 2i g d (\pi^0 F)
 \right] \,.
 \fe
 The equations of motion receive a bulk contribution, ${1\over e^2} d\star F =  i \star J^\text{EM}$, and a boundary contribution  on the defect $x=0$:
\ie
2ig \left( \pi^0\Big|_{x=0^+} - \pi^0\Big|_{x=0^-} \right) F= {i\over 2\pi } da \,.
\fe
 Combining with \eqref{pidiscont} and \eqref{aeom}, we find 
 \ie
 g={1\over 8\pi^2 f_\pi}\,.
 \fe  
Hence,  $\pi^0F\wedge F$ term in the pion Lagrangian \eqref{IRL} is necessary to match the non-invertible symmetry in the UV QCD Lagrangian.

Since the UV QCD Lagrangian is invariant  under gauging the magnetic one-form symmetry, so should the IR pion Lagrangian.
Indeed, upon gauging the $\mathbb{Z}_N^{(1)}$ magnetic one-form symmetry as in Appendix \ref{app:Maxwell}, the $\theta$-angle is shifted by $-2\pi p/N$, which can be undone by a $U(1)_\text{A3}$ rotation that shifts $\pi^0 \to \pi^0- 2\pi  f_\pi p/N$.  
The gauging argument rigorously proves that the non-invertible symmetry ${\cal D}_{p\over N}$ is a topological operator.

Let us compare our reasoning with the usual derivation in the literature. 
 The  $\pi^0F\wedge F$ term in the effective pion Lagrangian is conventionally argued using the ABJ anomaly. 
Since the fine structure constant is small, one can effectively treat the $U(1)_\text{EM}$ gauge field as a background gauge field, and interpret the ABJ anomaly as an 't Hooft anomaly between $U(1)_{\text{A}3}$ and $U(1)_\text{EM}$. 
The 't Hooft anomaly matching condition then determines the $\pi^0 F\wedge F$ coupling.  
The $\pi^0F\wedge F$ term can also be derived from the Wess-Zumino term in the chiral Lagrangian when coupled to the electromagnetic gauge field \cite{Witten:1983tw}. 
In this paper, we  provide an alternative derivation of the neutral pion decay from a matching condition for the non-invertible global symmetries for any finite and nonzero fine structure constant.

\section{Conclusion and Outlook}

In the past few years, there have been a lot of exciting developments on generalized global symmetries in high energy physics and condensed matter physics. 
In this paper, we identify some of the first examples of non-invertible global symmetries in Nature. 

In QED,   the continuous, invertible classical $U(1)_\text{A}$ symmetry   turns into a discrete, non-invertible global symmetry generated by the topological operators ${\cal D}_{p\over N}$, each labeled by a rational number $p/N \in \mathbb{Q}/\mathbb{Z}$. 
The non-invertible symmetry operator ${\cal D}_{p\over N}$ is a composition of the naive axial rotation with a rational angle $\alpha=2\pi p/N$,  together with a $\nu = p/N$ fractional quantum Hall state.

The operator ${\cal D}_{p\over N}$ acts invertibly on the fermions and the Wilson lines, but non-invertibly on the 't Hooft lines. 
It leads to selection rules on scattering amplitudes, which explain the familiar helicity conservation of electrons and positrons from a global symmetry principle.  

We similarly construct these non-invertible symmetries in QCD of the first generation in the massless limit and below the electroweak scale.  
The coupling $\pi^0 F\wedge F$ in the IR pion Lagrangian is necessary to match these non-invertible symmetries in the QCD Lagrangian. 
Therefore, the neutral pion decay $\pi^0\to \gamma\gamma$ is a direct consequence of the non-invertible global symmetries.

Our non-invertible global symmetries are only exact when the fermions are massless. Said differently, electrons and quarks are naturally massless in QED and QCD, respectively,  because of the non-invertible global symmetry.

There are several future directions:
\begin{itemize}
\item  What is the full non-invertible fusion algebra of ${\cal D}_{p\over N}$ and the condensation operator? 
\item How do we understand the spontaneous symmetry breaking of non-invertible symmetries? In QED, even though the non-invertible symmetry is discrete, it has infinitely many elements each labeled by a rational number $\mathbb{Q}/\mathbb{Z}$, which is dense in $U(1)$.  Is it possible to interpret $\pi^0$ as a Goldstone boson for this discrete but infinite non-invertible global symmetry? 
It is intriguing to speculate that this might be the reason why $\pi^0$ can be so light as a Goldstone boson, but also admits the non-derivative coupling $\pi^0 F\wedge F$. 
\item In this paper we consider ABJ anomalies where all the participating symmetries are $U(1)$. It would be interesting to extend our construction to non-abelian gauge groups. 
However, for simply connected  non-abelian gauge groups, there is no magnetic one-form symmetry, and our construction does not generalize to these cases straightforwardly. 
\item Can these non-invertible symmetries be gauged? If not, how do we understand their generalized 't Hooft anomalies? The dynamical consequences of similar non-invertible symmetries have been discussed in \cite{Choi:2021kmx,Choi:2022zal}.
\item If we interpret the pion field $\pi^0$ of  \eqref{IRL} as an axion, we immediately conclude that there are infinitely many non-invertible symmetries ${\cal D}_{p\over N}$ in the axion-Maxwell theory.\footnote{The higher group structure of axion gauge theory has been explored in \cite{Hidaka:2020izy,Brennan:2020ehu,Hidaka:2021mml,Hidaka:2021kkf,Kaya:2022edp}.}  
See \cite{Cordova:2022ieu} for applications  of non-invertible symmetries on axion physics.
\item In \cite{Roumpedakis:2022aik}, it was shown that the higher gauging of a higher-form symmetry leads to a non-invertible symmetry, generated by the  condensation operators. Using this construction, in addition to the symmetries discussed here, there are many other non-invertible symmetries from the higher gauging of higher-form symmetries (e.g., $\mathbb{Z}_6^{(1)}$) in the Standard Model.  We leave  these condensation operators for future investigations.
\end{itemize}

While the dynamical consequences from our non-invertible symmetries ${\cal D}_{p\over N}$ are all well-known (e.g., the helicity conservation and the pion decay), it is encouraging to see that these real world phenomena admit   alternative explanations in terms of  generalized global symmetries.  
It would be very exciting to explore other possible non-invertible global symmetries in the Standard Model and their dynamical applications.

 \section*{Acknowledgements}
 
 We are grateful to J.\ Albert, A.\ Cherman, C.\ Cordova, S.\ Dubovsky, M.\ Forslund, I.\ Halder, D.\ Harlow, J.\ Kaidi, Z.\ Komargodski, P.\ Meade, K.\ Ohmori, M.\ Reece,  N.\ Seiberg, S.\ Seifnashri, G.\ Sterman, and G.\ Zafrir   for useful discussions. 
 We thank J.\ Albert and Z.\ Komargodski for useful comments on a draft. 
 We would also like to thank C.\ Cordova and K.\ Ohmori for communications about their upcoming work \cite{Cordova:2022ieu} related to non-invertible symmetries from ABJ anomalies. 
 HTL is supported in part by a Croucher fellowship from the Croucher Foundation, the Packard Foundation and the Center for Theoretical Physics at MIT. 
The authors of this paper were ordered alphabetically. 

 \appendix

\section{Minimal $\mathbb{Z}_N$ TQFT ${\cal A}^{N,p}$}\label{app:minimalTQFT}

In this appendix, we give a lightening review of discrete one-form global symmetries in a spin TQFT and the minimal $\mathbb{Z}_N$ TQFT ${\cal A}^{N,p}$ in 2+1d.
The readers are referred to   \cite{Gaiotto:2014kfa,Hsin:2018vcg} for more details.  

In 2+1d, one-form global symmetries are generated by topological line operators. 
In particular, a $\mathbb{Z}_N^{(1)}$ one-form symmetry is generated by a topological symmetry line  $W$ obeying $W^N=1$. 
Two symmetry lines fuse according to the $\mathbb{Z}_N$ group multiplication law, i.e.,  $W^s\times W^{s'}=W^{s+s'}$, with $s,s'$ defined modulo $N$.  
In spin TQFTs, the 't Hooft anomaly of the $\mathbb{Z}_N$ one-form symmetry is parametrized by an integer $p\sim p+N$ \cite{Gaiotto:2014kfa,Gomis:2017ixy,Hsin:2018vcg}, which determines the topological spin of the symmetry lines:
\ie
h[W^s]=\frac{p s^2}{2N}\text{~~mod~~}\frac{1}{2}~.
\fe
The spin is defined modulo $\frac{1}{2}$ because we can always dress the symmetry line by  a transparent fermion line of spin $\frac{1}{2}$ in a spin TQFT. A line $V$ carries a one-form symmetry charge $q$   if it braids with $W^s$ with a phase $e^{2\pi iqs/N}$.

Interestingly, when $p$ and $N$ are coprime, these $N$ topological lines $\{W^s\}_{s=0,1,\cdots,N-1}$ themselves form a consistent 2+1d TQFT \cite{Hsin:2018vcg}. 
The coprime condition is needed so that the braiding is non-degenerate, i.e., the only line that braids trivially with all the other lines is the trivial line $1$. 
This theory is called the minimal $\mathbb{Z}_N$ TQFT, denoted by ${\cal A}^{N,p}$. Examples of minimal TQFTs include $\mathcal{A}^{N,1}=U(1)_N$ and $\mathcal{A}^{N,N-1}=SU(N)_1$.

It turns out every TQFT with such a $\mathbb{Z}_N^{(1)}$ one-form symmetry and anomaly factorizes into the minimal TQFT ${\cal A}^{N,p}$ and another decoupled TQFT consisting of the charge neutral lines \cite{Hsin:2018vcg}. As an example, consider $U(1)_N\times U(1)_N$ Chern-Simons theory with odd $N$
\ie
\frac{iN}{4\pi}a_1\wedge da_1+\frac{iN}{4\pi}a_2\wedge da_2~.
\fe
The topological Wilson lines $W_1=e^{i\oint (a_1+a_2)}$ and $W_2=e^{i\oint (a_1-a_2)}$ each has order $N$ and spin $\frac{2}{2N}\text{ mod }\frac{1}{2}$. Each of them generates a $\mathbb{Z}_N$ one-form symmetry of anomaly $p=2$. Since $p=2$ and $N$ are coprime, each of them also generates a minimal TQFT $\mathcal{A}^{N,2}$.  Hence the theory factorizes into the product of these two minimal TQFTs 
\ie
U(1)_N\times U(1)_N=\mathcal{A}^{N,2}\times\mathcal{A}^{N,2}~.
\fe 
The lines in the two descriptions are related as
\ie
e^{is_1\oint a_1}e^{is_2\oint a_2}=W_1^{(2^{-1})(s_1+s_2)} W_2^{(2^{-1})(s_1-s_2)}~,
\fe
where $2^{-1}$ is the multiplicative inverse of $2$ modulo $N$.

We can couple the minimal $\mathbb{Z}_N$ TQFT ${\cal A}^{N,p}$ to the background two-form gauge field $B$ for the $\mathbb{Z}_N^{(1)}$ one-form symmetry. The Lagrangian for the coupling is denoted by ${\cal A}^{N,p}[B]$.  
For example, when $p=1$, the Lagrangian is
\ie
\mathcal{A}^{N,1}[B]=\frac{iN}{4\pi} a\wedge da+\frac{iN}{2\pi}a\wedge B~.
\fe
Here $B$ is a flat $U(1)$ two-form gauge field whose holonomy is restricted to be $2\pi/N$ times an integer so that it's effectively a $\mathbb{Z}_N$ gauge field.

Because of the 't Hooft anomaly, the partition function of the theory is not invariant under the background one-form gauge transformation $B\rightarrow B+d\Lambda$. The anomalous gauge transformation can be canceled by extending the background gauge field $B$ to a 3+1d bulk $X_4$, with the classical action \cite{Hsin:2018vcg}
\ie\label{eq:One-form_SPT}
-\frac{ipN}{4\pi}\int_{X_4} B\wedge B~.
\fe
The classical action can be interpreted as a 3+1 $\mathbb{Z}_N^{(1)}$ one-form SPT phase \cite{Kapustin:2013uxa,Gaiotto:2014kfa,Thorngren:2015gtw}.

Instead of coupling the one-form symmetry to a background two-form gauge field $B$, we can activate the background using the field strength of a background $U(1)$ gauge field $A$. More specifically, we set $B=dA/N$ in the Lagrangian ${\cal A}^{N,p}[B]$. 
In the $p=1$ case,  $\mathcal{A}^{N,1}[dA/N]$  is the Lagrangian for the $\nu=1/N$ fractional quantum Hall state 
in  \eqref{FQH2}, which has a fractional Hall conductivity $\sigma_{xy}=-\frac{1}{N}$ described by the naive action \eqref{FQH1}. 
Similarly, the theory ${\cal A}^{N,p}[dA/N]$ has a fractional Hall conductivity $\sigma_{xy}=-\frac{p}{N}$ described by the naive Lagrangian
\ie
-\frac{ip}{4\pi N} A\wedge dA~.
\fe
Indeed, this naive Lagrangian cancels the anomaly inflow from the bulk \eqref{eq:One-form_SPT}  if we substitute $B=dA/N$
\ie
-\frac{ipN}{4\pi}\int_{X_4} B\wedge B=-\frac{ip}{4\pi N}\int_{X_4} d(A\wedge dA)=\frac{ip}{4\pi N}\int_{\partial X_4} A\wedge dA~.
\fe

The minimal $\mathbb{Z}_N$ TQFT $\mathcal{A}^{N,p}[B]$ can also be realized on the boundary of a 3+1d system \cite{Hsin:2018vcg}. Consider a 3+1d $\mathbb{Z}_N$ two-form gauge theory with the following action
\ie\label{bulkminimal}
\int_{X_4}\left( \frac{iN}{2\pi} b \wedge dc + \frac{iNk}{4\pi} b \wedge b+\frac{iN}{2\pi} b \wedge B\right)~,
\fe
where $k$ is the multiplicative inverse of $p$ modulo $N$. 
Here $c$ and $b$ are dynamical $U(1)$ one-form and two-form gauge fields, respectively. 
There is a zero-form and a one-form gauge symmetry
\ie
&b\rightarrow b+d\Lambda~,
\\
&c\rightarrow c-k\Lambda+d\lambda~,
\fe
where $\lambda$ and $\Lambda$ are the zero-form and one-form gauge parameters, respectively. 
The equation of motion for $b$ is
\ie\label{eq:EOM}
Nkb+Ndc+NB=0~.
\fe
Next, we integrate out $c$ to constrain $b$ to be a $\mathbb{Z}_N$ two-form gauge field, i.e., $N b = d \ell$ for some $U(1)$ one-form gauge field $\ell$. 
Then the equation of motion  of $b$ is equivalent to $b= - pB$ as $\mathbb{Z}_N$ gauge fields. 
Substituting it back to the action, we recover the classical action \eqref{eq:One-form_SPT}, where we have used the fact that $b,B$ are $\mathbb{Z}_N$ two-form gauge fields and dropped terms that are in $2\pi i\mathbb{Z}$.  This implies that the bulk theory is a $\mathbb{Z}_N$ one-form SPT phase with  trivial topological order.

Let us place \eqref{bulkminimal} on a  manifold with  boundary and impose Dirichlet boundary condition $b\vert=0$ which explicitly breaks the one-form gauge symmetry. Because of this, the line operator $V(\gamma)=\exp(i\oint_{\gamma} c)$ becomes gauge-invariant on the boundary. Its correlation function is
\ie
\left\langle V(\gamma)V(\gamma')\right\rangle=\exp\left(\frac{2\pi ik}{N}\ell(\gamma,\gamma')\right)~,
\fe
where $\text{link}(\gamma,\gamma')$ is the linking number between $\gamma$ and $\gamma'$. From the correlation function, we can read off the spin of $V$: $h[V]=\frac{k}{2N}$ mod $\frac{1}{2}$. Since $N$ and $k$ are coprime and $V$ obeys $V^N=1$, the $N$ lines generated by $V$ form a minimal $\mathbb{Z}_N$ TQFT $\mathcal{A}^{N,k}$ on the boundary. The theory $\mathcal{A}^{N,k}$ is the sames as $\mathcal{A}^{N,p}$ if we redefine the generator to be $V^p$. This a valid redefinition because $N$ and $p$ are coprime. The background gauge field $B$ couples to the line $V^p$ on the boundary. This is because $B$ couples to the surface operator $\exp(i\oint b)$ in the bulk and when $\exp(i\int_\Sigma b)$ ends on the boundary, the latter is equivalent to the line operator $V^p=\exp(ip\oint_{\partial\Sigma} c)$ using the equation of motion of $b$ \eqref{eq:EOM}. 

Combining all these facts, we conclude that the theory \eqref{bulkminimal} on a manifold with boundary is equivalent to a $\mathbb{Z}_N^{(1)}$ one-form SPT \eqref{eq:One-form_SPT} in the bulk and a  minimal TQFT $\mathcal{A}^{N,p}$ on the boundary described by the action
\ie
-\int_{X_4} \frac{ipN}{2\pi} B \wedge B + \int_{\partial X_4}\mathcal{A}^{N,p}[B]~.
\fe
This action is invariant under the background one-form gauge symmetry of $B$.

  \section{Gauging One-Form Symmetries in the Maxwell Theory}\label{app:Maxwell}

Gauging a discrete $q$-form symmetry on a codimension-$p$ manifold in spacetime  is known as the $p$-gauging  of a $q$-form symmetry \cite{Roumpedakis:2022aik}. 
The zero-gauging corresponds to the ordinary gauging of a higher-form symmetry. 
When $p>0$, the $p$-gauging does not change the bulk of the QFT, but generates a codimension-$p$ topological operator/defect, known as the condensation operator/defect, in the same QFT.

In Appendix \ref{app:zerogauging}, we discuss the ordinary gauging  of the magnetic one-form symmetry in the Maxwell theory, and identify two kinds of gaugings that only shift the $\theta$-angle while leaving the electric coupling invariant. 
When applied to QED, this shift in the $\theta$-angle can be undone by an axial rotation \eqref{chiralrotation}, and therefore QED is invariant under such gaugings.  
In Appendix \ref{app:onegauging}, we one-gauge the magnetic one-form symmetry to generate the magnetic condensation operators in the Maxwell theory. 
Since the magnetic one-form symmetry is not broken by the coupling to the electrically charged field (e.g., electrons),  our results here hold true for QED too. 
 
\subsection{Gauging Magnetic One-Form Symmetries and the $\theta$-Angle}\label{app:zerogauging}

Consider the free Maxwell theory with the complexified coupling constant $\tau = \frac{2\pi i}{e^2} + \frac{\theta}{2\pi}$. 
We denote this QFT by ${\cal Q}_\tau$. 
It has a $U(1)^{(1)}$ magnetic one-form symmetry generated by the charge $Q_m = \frac{1}{2\pi} \oint F \in \mathbb{Z}$ which is conserved due to the Bianchi identity $dF = 0$ \cite{Gaiotto:2014kfa}.
We can pick a subgroup $\mathbb{Z}_N^{(1)} \subset U(1)^{(1)}$ generated by $\exp \left( \frac{2\pi i}{N} Q_m \right)$, and couple the theory to a background gauge field for this $\mathbb{Z}_N^{(1)}$ subgroup which we denote as $B$.
The Lagrangian in the precense of the background gauge field $B$ is given by
\begin{equation} \label{Eq:Maxwell_Lag}
    \mathcal{L}_{\mathcal{Q}_\tau}[B] = \frac{1}{2e^2}F \wedge \star F + \frac{i\theta}{8\pi^2} F \wedge F
    + \frac{i}{2\pi} F \wedge B \,.
\end{equation}
Here $B$ is a flat $U(1)$ two-form gauge field whose holonomy is restricted to be $2\pi/N$ times an integer so that it's effectively a $\mathbb{Z}_N$ gauge field.
The QFT $\mathcal{Q}_\tau$ is defined modulo counterterms that are independent of $B$, such as gravitational counterterms.

Given a $\mathbb{Z}_N^{(1)}$ one-form global symmetry and an integer $k$ modulo $N$ obeying gcd$(k,N)=1$, we define the following operations \cite{Gaiotto:2014kfa,Bhardwaj:2020ymp,Choi:2022zal}\footnote{The operations $S^{(k)}$, $T^{(k)}$, and $C$ are not to be confused with the electromagnetic dualities of the Maxwell theory. We don't use the electromagnetic dualities in this work.}
\begin{align}
\begin{split}
    S^{(k)} &: \quad \mathcal{L}_{S^{(k)}\mathcal{Q}_\tau}[B] = \mathcal{L}_{\mathcal{Q}_\tau}[b]
    +\frac{iN}{2\pi} b \wedge dc 
    + \frac{iNk}{2\pi} b \wedge B \,, \\
    T ^{(k)} &: \quad \mathcal{L}_{T^{(k)}\mathcal{Q}_\tau}[B] = \mathcal{L}_{\mathcal{Q}_\tau}[B]
    +\frac{iNk}{4\pi} B \wedge B \,, \\
    C &: \quad \mathcal{L}_{C\mathcal{Q}_\tau}[B] = \mathcal{L}_{\mathcal{Q}_\tau}[-B] \,.
\end{split}
\end{align}
We adopt the convention where the upper case letters denote background gauge fields, and the lower case letters denote dynamical gauge fields. 
The only exception to this convention is that we use $A$ for the dynamical $U(1)$ gauge field for the Maxwell theory, which is to be distinguished from the dynamical $U(1)$ gauge field $a$ on the non-invertible operator in the main text. 
Here $b$ is a $U(1)$ two-form gauge field, and $c$ is a $U(1)$ one-form gauge field whose equation of motion constrains the former to be $\mathbb{Z}_N$-valued.

The $S^{(k)}$ operation corresponds to gauging the $\mathbb{Z}_N^{(1)}$ subgroup of the magnetic one-form symmetry.
Here the integer $k$ specifies a generator of the quantum $\mathbb{Z}_N^{(1)}$ one-form symmetry that we obtain after gauging to be $\exp \left( ik\oint b \right)$.
The $T^{(k)}$ operation corresponds to stacking an SPT phase for the $\mathbb{Z}_N^{(1)}$ subgroup given by the partition function $\exp \left( \frac{iNk}{4\pi}\int B \wedge B \right)$.\footnote{When $N$ is odd, the $T^{(k)}$ operation is well-defined only if the spacetime manifold is spin, which we  assume throughout this paper.}
When $k=1$, we will simply write $S \equiv S^{(1)}$ and $T\equiv T^{(1)}$.
The $C$ operation flips the sign of the background gauge field $B$ (which can be loosely referred to as a ``charge conjugation''). 
It becomes trivial if $N=2$.

These operations satisfy the following relations:
\begin{equation} \label{Eq:SL2Z}
    \left( S^{(k)} \right)^2 = C \,, \quad
    \left( S^{(k)} T^{(k)} \right)^3 = Y^{(k)} \,, \quad
   ( T^{(k)})^N = 1 \,, \quad
    C^2 = 1 \,.
\end{equation}
Here, $Y^{(k)}$ is an invertible theory whose Lagrangian is given by
\begin{equation}
    \mathcal{L}_{Y^{(k)}} = \frac{iNk}{4\pi} b \wedge b
        + \frac{iN}{2\pi} b \wedge dc \,.
\end{equation}
Upon integrating out $b$ and $c$, $Y^{(k)}$ becomes a gravitational counterterm. 
Its partition function on a closed four-manifold is independent of $k$ and  is given by $e^{2\pi i \sigma}$ (modulo the Euler counterterm), where $\sigma$ is the signature of the four-manifold.  
Here we assume that the integral cohomology has no torsion. 
This phase is nontrivial on general non-spin manifolds but becomes trivial on spin manifolds (see, for example, \cite{Gaiotto:2014kfa} for more discussions on this invertible field theory). 

The action of $S^{(k)}$, $T^{(k)}$, and $C$ realizes the group $SL(2,\mathbb{Z}_N)^{(k)}$ projectively.
The superscript in $SL(2,\mathbb{Z}_N)^{(k)}$ is there to keep track of the fact that there are different $SL(2,\mathbb{Z}_N)$ operations corresponding to different choices of $k$.\footnote{Note that the intersections between $SL(2,\mathbb{Z}_N)^{(k)}$ with different values of $k$ are not empty. 
For instance, $S^{(k)} \in SL(2,\mathbb{Z}_N)^{(k)}$ but $S^{(k)} \not\in SL(2,\mathbb{Z}_N)^{(k')}$ if $k \neq k'$ mod $N$. On the other hand, $T$ and $C$ are elements of $SL(2,\mathbb{Z}_N)^{(k)}$ for any $k$.}
The groups $SL(2,\mathbb{Z}_N)^{(k)}$ act projectively on the Maxwell theories ${\cal Q}_\tau$ where the projective phase is determined by the invertible theory $Y^{(k)}$. 

Now, we will discuss some composite operations  labeled by group elements in $SL(2,\mathbb{Z}_N)^{(k)}$, which leave the coupling constant $e$ intact while shifting $\theta$ by a certain angle. 
In QED, we can further undo the shift in the $\theta$-angle by an axial rotation \eqref{chiralrotation}, and return to the original QED.

We start with the operation $CT^{(k)} S^{(k)} T^{(k)} \in SL(2,\mathbb{Z}_N)^{(k)}$. 
The $CT^{(k)} S^{(k)} T^{(k)}$ gauging of the Maxwell theory is given by the Lagrangian
\begin{align}\label{LCTSTQ}
\begin{split}
    \mathcal{L}_{CT^{(k)} S^{(k)} T^{(k)} {\cal Q}_\tau}[B] &=
    \frac{1}{2e^2} F \wedge \star F + \frac{i\theta}{8\pi^2} F \wedge F + \frac{i}{2\pi} F \wedge b \\
    & \quad + \frac{iN}{2\pi} b \wedge dc
    + \frac{iNk}{4\pi} (b-B) \wedge (b-B) \,.
\end{split}
\end{align}
The gauge transformations of the dynamical gauge fields are given by
\begin{align}
    \begin{split}
        b &\sim b + d\Lambda \,,\\
        c &\sim c - k\Lambda+ d\lambda \,,\\
    \end{split}
    \end{align}
where $\Lambda$ is a one-form gauge parameter and $\lambda$ is a zero-form gauge parameter.  

We then integrate out $c$, which enforces $b$ to be a $\mathbb{Z}_N$ two-form gauge field, i.e., $N b = d \ell$ for some $U(1)$ one-form gauge field $\ell$. 
Let $p$ be an integer satisfying $pk = 1$ mod $N$.
Such a $p$ uniquely exists modulo $N$ since $\text{gcd}(k,N)=1$. 
We can then rewrite the Lagrangian as
\begin{equation}
    \frac{1}{2e^2} F \wedge \star F + \frac{i}{8\pi^2} \left( \theta - \frac{2\pi p}{N} \right) F \wedge F + \frac{i}{2\pi} F \wedge B
 +  {iNk\over 4\pi}\left ( b - B+\frac{p}{N}F  \right)^2 
     \,.
\end{equation}
where we have used the fact that $b,B$ are $\mathbb{Z}_N$ two-form gauge fields and dropped terms that are in $2\pi i\mathbb{Z}$. 
Note that every term in the last parentheses is  a properly normalized $\mathbb{Z}_N$ two-form gauge field, 
and the equation of motion for $b$ sets $b= B- {p\over N} F$.
Alternatively, we can shift $b$ to rewrite it as ${iNk\over 4\pi} b\wedge b$, which upon integrating out $b$ leads to a gravitational counterterm that is independent of $B$.

Comparing with \eqref{Eq:Maxwell_Lag}, we conclude that up to counterterms that are independent of $B$,
\begin{equation}
   CT^{(k)} S^{(k)} T^{(k)} \mathcal{Q}_\tau = \mathcal{Q}_{\tau - \frac{p}{N}} \,.
\end{equation}
That is, the $CT^{(k)} S^{(k)} T^{(k)}$ gauging operation simply shifts the $\theta$-angle by $-2\pi p/N$.
Similarly, the inverse operation gives $(T^{(k)})^{-1} S^{(k)} (T^{(k)})^{-1} \mathcal{Q}_\tau = \mathcal{Q}_{\tau + \frac{p}{N}}$.

Next, we consider another operation given by the sequence $CST^{-p} S \in SL(2,\mathbb{Z}_N)^{(1)}$ (recall $S \equiv S^{(1)}$ and $T\equiv T^{(1)}$).
Here, we will not assume $\text{gcd}(p,N) =1$ and $p$ can be any arbitrary integer defined modulo $N$.
The Lagrangian in this case is given by
\ie
\mathcal{L}_{CST^{-p}S\mathcal{Q}_\tau}[B]=&\,
    \frac{1}{2e^2}F \wedge \star F + \frac{i\theta}{8\pi^2} F \wedge F+\frac{i}{2\pi} F\wedge b_1+\frac{iN}{2\pi}b_1\wedge dc_1
    \\&\,+\frac{iN}{2\pi}b_1\wedge b_2-\frac{ipN}{4\pi}b_2\wedge b_2+\frac{iN}{2\pi}b_2\wedge dc_2-\frac{iN}{2\pi}b_2\wedge B~.
\fe
The gauge transformations are
\ie
&b_1\sim b_1+d\Lambda_1~,
\\
&c_1\sim c_1-\Lambda_2 + d\lambda_1~,
\\
&b_2\sim b_2+d\Lambda_2~,
\\
&c_2\sim c_2-\Lambda_1 + p\Lambda_2 + d\lambda_2~.
\fe
As before, we can first integrate out $c_1$ and $c_2$ which forces $b_1$ and $b_2$ to be $\mathbb{Z}_N$ two-form gauge fields.
That is, we have $Nb_1 = d\ell_1$ and $Nb_2 = d\ell_2$ for some $U(1)$ one-form gauge fields $\ell_1$ and $\ell_2$.
The Lagrangian can be rewritten as
\ie
    \frac{1}{2e^2}F \wedge \star F + &\frac{i}{8\pi^2}
    \left(\theta - \frac{2\pi p}{N} \right)
    F \wedge F+\frac{i}{2\pi}F\wedge B \\
    &+\frac{iN}{2\pi} \left( b_2 + \frac{1}{N}F \right) \wedge \left( b_1 + \frac{p}{N}F - B \right)
    -\frac{ipN}{4\pi} \left( b_2 + \frac{1}{N}F   \right)^2
    ~.
\fe
In the last two terms, $b_1$ and $b_2$ can be appropriately shifted to remove the dependence on $F$ and $B$, as all the terms are properly normalized $\mathbb{Z}_N$ two-form gauge fields.
Then, the contribution from these terms is a gravitational counterterm which is independent of $B$.
Modulo the counterterms independent of $B$, we thus obtain
\ie\label{eq:STS}
CST^{-p} S\mathcal{Q}_\tau=\mathcal{Q}_{\tau-\frac{p}{N}}~.
\fe
That is, the $CST^{-p}S$ operation simply shifts the $\theta$-angle of the Maxwell theory by $-2\pi p/N$.
This is similar to the previous $CT^{(k)} S^{(k)} T^{(k)}$ gauging operation.
However, for the case of the $CST^{-p}S$ gauging operation, we see that this holds even without the condition $\text{gcd}(p,N)=1$.
One nice  feature of this gauging is that the set of elements $CST^{-p}S \in SL(2,\mathbb{Z}_N)^{(1)}$ for different values of $p$ forms a $\mathbb{Z}_N$ subgroup of the modular group $SL(2,\mathbb{Z}_N)^{(1)}$. Indeed, we have $(CST^{-p_1}S)(CST^{-p_2}S) = CST^{-p_1 - p_2}S$.

The results in this appendix are still applicable even if there are electrically charged matter fields coupled to the Maxwell theory (such as QED discussed in the main text), since such matter fields do not affect the magnetic one-form symmetry and its gauging.

\subsection{Condensation Operators from One-Gauging}\label{app:onegauging}

Here we discuss  condensation operators/defects from one-gauging  a $\mathbb{Z}_N^{(1)}$ subgroup of the magnetic one-form symmetry along a codimension-one submanifold $M$ in  spacetime.  
The condensation operators/defects from  one-gauging  the electric one-form symmetry in the Maxwell theory have been discussed in  \cite{Roumpedakis:2022aik,Choi:2022zal}.

If we denote the $\mathbb{Z}_N^{(1)}$ magnetic one-form symmetry operator supported on a two-cycle $\Sigma \subset M$ as $\eta(\Sigma) = \exp \left(\frac{i}{N} \oint_\Sigma F \right)$, the condensation operator can be expressed as \cite{Roumpedakis:2022aik,Choi:2022zal} (see also \cite{Choi:2021kmx,Kaidi:2021xfk})
\begin{equation} \label{eq:cond_op}
    \mathcal{C}_\epsilon (M) = \frac{1}{|H^0(M,\mathbb{Z}_N)|}
    \sum_{\Sigma \in H_2(M,\mathbb{Z}_N)} \epsilon(M,\Sigma) \eta(\Sigma)
\end{equation}
where $\epsilon(M,\Sigma)$ is a discrete torsion phase  on $M$ for the one-gauging.  
Since the one-gauging of a one-form symmetry is similar to the ordinary gauging of a zero-form symmetry of  a 2+1d spin QFT on $M$, the discrete torsion phases are classified by $\text{Hom}(\text{Tors} \,\Omega_3^{\text{Spin}}(B\mathbb{Z}_N),U(1))$. 
This group is given by (see, for example, \cite{Cordova:2017vab})
\ie \label{eq:spin_bordism}
\text{Hom}(\text{Tors} \,\Omega_3^{\text{Spin}}(B\mathbb{Z}_N),U(1))=
\begin{cases}
&\mathbb{Z}_N\,,~~~~~~~~~~\,~~\text{if}~N=1~\text{mod}~2\,,\\
&\mathbb{Z}_{2N}\times \mathbb{Z}_2\,,~~\text{if}~N=0~\text{mod}~4\,,\\
&\mathbb{Z}_{4N}\,,~~~~~~~~~~~\text{if}~N=2~\text{mod}~4\,.
\end{cases}
\fe
For instance, for $N=2$, this group is $\mathbb{Z}_8$ \cite{2010PhRvB..81m4509F}, and we expect that there are eight different condensation operators.

Out of all possible condensation operators, we will focus only on the following:
\begin{equation} \label{eq:cond_lag}
    \mathcal{C}_K (M) \equiv \exp \left[i\oint_M
    \left(
    \frac{N}{2\pi} a \wedge d\tilde{a} + \frac{K}{4\pi} a\wedge da + \frac{1}{2\pi} a \wedge dA 
    \right)
\right] \,. 
\end{equation}
Here, $a$ and $\tilde{a}$ are dynamical $U(1)$ gauge fields living on the submanifold $M$, and $K$ is an integer whose allowed values and  periodicity will be explained below. 
We now relate the above expression with \eqref{eq:cond_op}. 
The equation of motion of $\tilde a$ from the term $\frac{N}{2\pi} a \wedge d\tilde{a}$ forces $a$ to effectively become a $\mathbb{Z}_N$ gauge field.
Then, the remaining path integral over $a$ collapses to the finite sum over $H^1(M,\mathbb{Z}_N)$, which in turn is equivalent to the sum over $H_2(M,\mathbb{Z}_N)$ in  \eqref{eq:cond_op} through the Poincar\'e duality.
More specifically, we have $\exp \left( \frac{i}{2\pi} \oint_M a \wedge dA \right) = \exp \left( \frac{i}{N} \oint_\Sigma dA \right) = \eta (\Sigma)$ where $\Sigma = \text{PD}(a)$ is the Poinca\'e dual of $a$ in $M$, and the path integral  over $a$ has turned into a sum over $\Sigma \in H_2(M,\mathbb{Z}_
N)$.
Finally, the middle term $\frac{K}{4\pi} a \wedge da$ corresponds to a particular choice of the discrete torsion $\epsilon$.

To understand the discrete torsion given by $\frac{K}{4\pi} a \wedge da$ better, first recall the following presentation of the 2+1d twisted $\mathbb{Z}_N$ gauge theory at level $K$ in terms of the $U(1)$ gauge fields $a$ and $\tilde{a}$ \cite{Maldacena:2001ss,Banks:2010zn,Kapustin:2014gua},
\begin{equation} \label{Eq:ZN_level_K}
    (\mathcal{Z}_N)_K \colon \quad
        \frac{iK}{4\pi} a \wedge da + \frac{iN}{2\pi} a \wedge d \tilde{a}\,.
\end{equation}
Here, we follow the convention of \cite{Hsin:2018vcg}.
Comparing with Eq. \eqref{eq:cond_lag}, we see that the condensation operator $\mathcal{C}_K(M)$ is simply
\begin{equation}
	\mathcal{C}_K(M) = \exp \left[ 
		\oint_M \left(  \mathcal{Z}_N \right)_K [dA/N]
	\right] \,.
\end{equation}
That is, the condensation operator $\mathcal{C}_K (M)$ corresponds to having a twisted $\mathbb{Z}_N$ gauge theory $(\mathcal{Z}_N)_K$ at level $K$ along the submanifold $M$ where a background gauge field configuration for the $\mathbb{Z}_N^{(1)}$ one-form symmetry   is activated by coupling to the bulk gauge field $dA/N$. This is analogous to the relation between the operator $\mathcal{D}_{\frac{p}{N}}$ \eqref{Eq:noninv_defect_kN} and the minimal TQFT $\mathcal{A}^{N,p}$.

This interpretation of the condensation operators \eqref{eq:cond_lag} immediately tells us what are the allowed values of $K$ in $\mathcal{C}_K(M)$ from those of the $(\mathcal{Z}_N)_K$ gauge theory.
On spin manifolds, $K$ can be any integer and the periodicity is $K\sim K+2N$ if $N$ is even and $K\sim K+N$ if $N$ is odd.\footnote{On non-spin manifolds,  $K$ is an even integer with periodicity $K\sim K+2N$, which labels a class in   $H^3(B\mathbb{Z}_N,U(1))\cong \mathbb{Z}_N$. In this case, $(\mathcal{Z}_N)_K$ is a Dijkgraaf-Witten gauge theory \cite{Dijkgraaf:1989pz}. } 
Comparing with   \eqref{eq:spin_bordism}, we see that this class of condensation operators given by \eqref{eq:cond_lag} captures only a half of all possible condensation operators if $N$ is even, and it captures all possible condensation operators if $N$ is odd.

In the derivation of some of the non-invertible fusion rules in Section \ref{sec:algebra}, we have in particular encountered the condensation operator $\mathcal{C}_K$ at level $K=N$. 
Note that when $N$ is odd, since the periodicity of the level $K$ is $K \sim K+N$,  we have $\mathcal{C}_N = \mathcal{C}_0$. 
We will denote this condensation operator simply as $\mathcal{C}$:\footnote{The condensation operators ${\cal C}_k(M)$ from one-gauging a $\mathbb{Z}_N^{(1)}$ one-form symmetry in a  3+1d non-spin QFT were analyzed in \cite{Choi:2022zal}. In the non-spin case, the discrete torsion for one-gauging is given by a class in $ H^3(B\mathbb{Z}_N,U(1))\cong \mathbb{Z}_N$ labeled by an integer $k$ modulo $N$. For even $N$, the operator ${\cal C}_{k= N/2}$ in \cite{Choi:2022zal} corresponds to the operator ${\cal C}_{K=N}(M)$ here in the spin case. }
\begin{equation}
	\mathcal{C}(M) \equiv 
	\begin{cases}
	\mathcal{C}_N(M) \,,~~~\text{for even}~N\,,\\
		{\cal C}_0(M) \,,~~~~\text{for odd}~N\,.
		\end{cases}
\end{equation}

Since the condensation operator is made of lower-dimensional surfaces \eqref{eq:cond_op}, it acts trivially on all local operators. 
On the other hand, as we sweep the condensation operator ${\cal C}_\epsilon(M)$ topologically past an 't Hooft line, it will leave behind a line operator that only lives on  ${\cal C}_\epsilon(M)$.  
This line can be thought of as the quantum (dual) symmetry of the one-gauging on $M$, which is called the higher quantum symmetry line in \cite{Roumpedakis:2022aik}.

Similar to before, these definitions of the condensation operators are valid even in the presence of electrically charged matter fields, since their presence doesn't break the magnetic one-form symmetry.

\section{Non-invertible Fusion Rules from Discrete Gauging}\label{app:fusion}
  
In the appendix, we derive the fusion rule of various non-invertible defects using the definition of gauging in half of the spacetime.
  
\subsection{Non-invertible Symmetries from the $CT^{(k)} S^{(k)} T^{(k)}$ Gauging}

In Section \ref{sec:gauging}, the operator $\mathcal{D}_{\frac{p}{N}}$ is defined by applying the $CT^{(k)} S^{(k)} T^{(k)}$ operation to the half-spacetime $x>0$ with the Dirichlet boundary condition imposed on $b$, where $pk = 1$ mod $N$.
The orientation reversal $\mathcal{D}^\dagger_{\frac{p}{N}}$ corresponds to applying the same operation to the $x<0$ region instead, again with the Dirichlet boundary condition.

Consider the parallel fusion $\mathcal{D}_{\frac{p}{N}} \times \mathcal{D}^\dagger_{\frac{p}{N}}$.
The special case $p=1$ was worked out in the main text using the worldvolume Lagrangian description of the operators.
Here we reproduce the same fusion rule by employing the description in terms of gauging in half of spacetime. 
From this picture, the fusion $\mathcal{D}_{\frac{p}{N}} \times \mathcal{D}^\dagger_{\frac{p}{N}}$ corresponds to applying the $CT^{(k)} S^{(k)} T^{(k)}$ operation in a thin slab $0<x<L$ where $L \to 0$.
We have the Dirichlet boundary condition $b|_{x=0} = b|_{x=L} =0$.
The Lagrangian inside the slab is given by\footnote{We set $B=0$ so that there are no additional magnetic one-form symmetry operators inserted.  }
\begin{equation}
    \mathcal{L}_{\text{QED}}[\Psi,\bar{\Psi},A] + \frac{i}{2\pi} F \wedge b
     + \frac{iN}{2\pi} b \wedge dc
    + \frac{iNk}{4\pi} b \wedge b
\end{equation}
whereas outside of the slab we simply have the Lagrangian $\mathcal{L}_{\text{QED}}[\Psi,\bar{\Psi},A]$.

In \cite{Kaidi:2021xfk,Choi:2021kmx}, it was shown that such a gauging of the one-form symmetry inside a thin slab with the Dirichlet boundary condition reduces to gauging on the codimension-one locus $x=0$.
The two-form $\mathbb{Z}_N$ gauge field $b$ reduces to the one-form $\mathbb{Z}_N$ gauge field $a$ living on the locus $x=0$, and the $b\wedge b$ term reduces to\footnote{This map $\text{Hom}(\text{Tors} \,\Omega_4^{\text{Spin}}(B^2 \mathbb{Z}_N),U(1)) \cong \mathbb{Z}_N\rightarrow \text{Hom}(\text{Tors} \,\Omega_3^{\text{Spin}}(B\mathbb{Z}_N),U(1))$ generalizes the similar map discussed in \cite{Roumpedakis:2022aik,Choi:2022zal} for the bosonic SPTs to the case of fermionic SPTs. For odd $N$, this is a trivial map, and for even $N$ it maps to an order 2 element.}
\begin{equation}
    \frac{iNk}{4\pi} b \wedge b \mapsto {iNk\over 4\pi} a\wedge da
\end{equation}
as was shown in \cite{Kaidi:2021xfk}.
Thus, we obtain the fusion rule
\begin{equation}
    \mathcal{D}_{\frac{p}{N}}(M) \times \mathcal{D}^\dagger_{\frac{p}{N}}(M) ={\cal C}(M)~.
\end{equation}
Note that $p$ and $N$ cannot both be even because $\text{gcd}(N,p)=1$. The special case $p=1$ (and thus $k=1$) reproduces \eqref{DDbar}.

Next, we will reproduce the fusion rule $\mathcal{D}_{\frac{1}{N}} \times \mathcal{D}_{\frac{1}{N}} = \mathcal{A}^{N,2} \,\mathcal{D}_{\frac{2}{N}}$ for the case of odd $N$ using the gauging picture.
The operator $\mathcal{D}_{\frac{1}{N}} \times \mathcal{D}_{\frac{1}{N}}$ is given by applying the $CT S T$ operation twice in the half-spacetime $x>0$ with the Dirichlet boundary condition.
The corresponding Lagrangian is 
\ie \label{eq:DxD}
&    \int_{x<0} \mathcal{L}_\text{QED}[\Psi,\bar{\Psi},A] 
    +{4\pi i \over 2N} \oint_{x=0} \star j^\text{A}+ \int_{x\geq 0} \left(
    \mathcal{L}_\text{QED}[\Psi,\bar{\Psi},A] + \frac{2i}{4\pi N} F \wedge F\right)\\ 
  &  + \int_{x\geq 0} \left(  \frac{i}{2\pi} b_1  \wedge F+\frac{iN}{2\pi} b_1 \wedge dc_1+\frac{iN}{4\pi} b_1 \wedge b_1-\frac{iN}{2\pi} b_1 \wedge b_2
    + \frac{iN}{2\pi} b_2 \wedge dc_2 + \frac{2iN}{4\pi} b_2 \wedge b_2
    \right) \,.
\fe
where $b_1|_{x=0} = b_2|_{x=0} =0$.
Let $b \equiv b_1$, $\tilde{b} \equiv b_1 -2 b_2$, $c \equiv c_1+2^{-1}c_2$, and $\tilde{c} \equiv -2^{-1}c_2$.
Here, $2^{-1}$ denotes the integer which is the inverse of 2 mod $N$.
Such an integer uniquely exists modulo $N$ since $N$ is odd.
In terms of the redefined fields, the Lagrangian \eqref{eq:DxD} becomes 
\ie \label{eq:DxD2}
&    \int_{x<0} \mathcal{L}_\text{QED}[\Psi,\bar{\Psi},A] 
    +{4\pi i \over 2N} \oint_{x=0} \star j^\text{A}\\
  &  + \int_{x\geq 0} \left(
    \mathcal{L}_\text{QED}[\Psi,\bar{\Psi},A] + \frac{2i}{4\pi N} F \wedge F + \frac{i}{2\pi} b \wedge F 
    +\frac{iN}{2\pi} b \wedge dc + \frac{iN}{4\pi} 2^{-1} b \wedge b
    \right)\\
  &  + \int_{x\geq 0} \left(\frac{iN}{2\pi} \tilde{b} \wedge d\tilde{c} + \frac{iN}{4\pi}2^{-1} \tilde{b} \wedge \tilde{b}
    \right) \,.
\fe
The first two lines give us the operator $\mathcal{D}_{\frac{2}{N}}$, whereas the last line corresponds to a decoupled minimal TQFT $\mathcal{A}^{N,2}$ at $x=0$ which does not couple to any of the QED fields.
Thus, we obtain the desired fusion rule 
\ie
\mathcal{D}_{\frac{1}{N}} \times \mathcal{D}_{\frac{1}{N}} = \mathcal{A}^{N,2} \,\mathcal{D}_{\frac{2}{N}}\,.
\fe

\subsection{Non-invertible Symmetries from the $CST^{-p}S$ Gauging}

In appendix  \ref{app:zerogauging}, we discuss a different gauging $CST^{-p}S$, which also leaves QED invariant when combined with an axial rotation. Hence, similar to the way we define $\mathcal{D}_{\frac{p}{N}}$ in Section \ref{sec:gauging}, we can define a different set of non-invertible symmetry $\tilde{\mathcal{D}}_{p,N}$ by performing a $CST^{-p}S$ gauging in half of the spacetime. The resulting topological operators $\tilde {\cal D}_{p,N}$ are not minimal in the sense that it can be decomposed into ${\cal D}_{p\over N}$ times a decoupled 2+1d TQFT when $\text{gcd}(N,p)=1$. 
However, the non-invertible fusion algebras of $\tilde {\cal D}_{p,N}$ are easier to compute, which we determine below.

Similar to Section \ref{sec:gauging}, the non-invertible symmetry $\tilde{\mathcal{D}}_{p,N}$ is defined by composing the following two actions in half of spacetime $x\ge 0$: (1) $CST^{-p}S$ gauging the $U(1)^{(1)}$ magnetic one-form symmetry, and (2) a change of variables by an axial rotation $\Psi\to \exp({2\pi i p \gamma_5\over 2N})\Psi$. On the boundary, we impose Dirichlet boundary condition $b_1\vert_{x=0}=b_2\vert_{x=0}=0$.  It is described by the corresponding action
\ie
&\int_{x<0} \mathcal{L}_\text{QED}[\Psi,\bar{\Psi},A] 
    +{2\pi ip \over 2N} \oint_{x=0} \star j^\text{A}+\int_{x\geq 0} 
    \left(\mathcal{L}_\text{QED}[\Psi,\bar{\Psi},A]+ \frac{ip}{4\pi N} F \wedge F\right)\\
  &  + \int_{x\geq 0} \left(
    \frac{i}{2\pi} F\wedge b_1+\frac{iN}{2\pi}b_1\wedge dc_1
    +\frac{iN}{2\pi}b_1\wedge b_2-\frac{ipN}{4\pi}b_2\wedge b_2+\frac{iN}{2\pi}b_2\wedge dc_2\right)~.
\fe
Following a similar analysis in  Appendix \ref{app:minimalTQFT}, we find that the line operators $V_1(\gamma)=\exp(i\oint_\gamma c_1)$ and $V_2(\gamma)=\exp(i\oint_\gamma c_2)$ become gauge-invariant  on the boundary
\begin{equation}
\begin{aligned}
&\langle V_1(\gamma) V_1(\gamma')\rangle=1~,
\\
&\langle V_1(\gamma) V_2(\gamma')\rangle=\exp\left(\frac{2\pi i}{N}\text{link}(\gamma,\gamma')\right)~,
\\
&\langle V_2(\gamma) V_2(\gamma')\rangle=\exp\left(-\frac{2\pi ip}{N}\text{link}(\gamma,\gamma')\right)~.
\end{aligned}
\end{equation}
where $\text{link}(\gamma,\gamma')$ is the linking number between $\gamma$ and $\gamma'$. They form a $(\mathcal{Z}_{N})_{-pN}$ gauge theory on the boundary. Using the equations of motion: $b_1=-d(pc_1+c_2)$, we find that the field strength $F=dA$ couples to the line operator $V_1^pV_2$ on the boundary. Combining all these, the non-invertible symmetry operator  takes the form 
\ie
\tilde{\mathcal{D}}_{p,N}=\exp\left[\oint_M\left(\frac{2\pi i p}{2N}\star j^\text{A}+\frac{iN}{2\pi}a_1 \wedge da_2-\frac{ipN}{4\pi}a_1\wedge da_1+\frac{i}{2\pi}a_2\wedge dA\right)\right]~,
\fe
where we have redefined $a_1=c_1$ and $a_2=pc_1+c_2$.
We can interpret the non-invertible symmetry $\tilde{\mathcal{D}}_{p,N}$ as a composition of the naive axial rotation and a  twisted $\mathbb{Z}_N$ gauge theory $(\mathcal{Z}_N)_{-pN}$. Unlike $\mathcal{D}_{\frac{p}{N}}$, the non-invertible symmetry $\tilde{\mathcal{D}}_{p,N}$ is defined for any $N$ and $p$ even if $p$ and $N$ are not coprime. When $p=0$, $\tilde{\mathcal{D}}_{0,N}$ becomes the condensation defect $\mathcal{C}_0$ associated to the $\mathbb{Z}_N^{(1)}$ magnetic one-form symmetry defined in Appendix \ref{app:Maxwell}.

When $N$ and $p$ are coprime, $(\mathcal{Z}_N)_{-pN}$ Chern-Simons gauge theory factorizes into two minimal TQFTs $\mathcal{A}^{N,p}$ and $\mathcal{A}^{N,-p}$. The first one is generated by $\exp(i\oint a_2)$ and the second one is generated by $\exp(i\oint (pa_1-a_2))$. Using this factorization property, we can relate $\tilde {\cal D}_{p, N}$ to the minimal non-invertible symmetry ${\cal D}_{p\over N}$ in \eqref{Eq:noninv_defect_kN}:
\ie
\tilde{\mathcal{D}}_{p, N}=\mathcal{D}_{\frac{p}{N}}\exp\left[\oint_M\mathcal{A}^{N,-p}\right]~,\quad \text{ when }\quad\text{gcd}(N,p)=1~.
\fe
Recall that $\mathcal{D}_{\frac{p}{N}}$ is defined using the $CT^{(k)} S^{(k)} T^{(k)}$ gauging.

It is straightforward to determine the fusion rule of $\tilde{\mathcal{D}}_{p,N}$. We have
\ie
\tilde{\mathcal{D}}_{p,N}\times \tilde{\mathcal{D}}_{p',N}=&\exp\left[\oint_M\left(\frac{2\pi i p}{2N}\star j^\text{A}+\frac{iN}{2\pi}a_1\wedge da_2-\frac{ipN}{4\pi}a_1\wedge da_1+\frac{i}{2\pi}a_2\wedge dA\right)\right]
\\
&\exp\left[\oint_M\left(\frac{2\pi i p'}{2N}\star j^\text{A}+\frac{iN}{2\pi}a_1'\wedge da_2'-\frac{ip'N}{4\pi}a_1'\wedge da_1'+\frac{i}{2\pi}a_2' \wedge dA\right)\right]~.
\fe
From the expression, it is clear that the fusion is commutative $\tilde{\mathcal{D}}_{p,N}\times \tilde{\mathcal{D}}_{p',N}=\tilde{\mathcal{D}}_{p',N}\times \tilde{\mathcal{D}}_{p,N}$.
Let $\bar{a}_1\equiv a_1$, $\bar{a}_2\equiv a_2+a_2'$, $\tilde a_1\equiv a_1'-a_1$ and $\tilde a_2\equiv a_2'-p'a_1$. In terms of the redefined fields, the defect becomes
\ie
~&\tilde{\mathcal{D}}_{p, N}\times \tilde{\mathcal{D}}_{p',N}=\exp\left[\oint_M\left(\frac{iN}{2\pi}\tilde a_1\wedge d\tilde a_2-\frac{ip'N}{4\pi}\tilde a_1\wedge d\tilde a_1\right)\right]
\\
&\times \exp\left[\oint_M\left(\frac{2\pi i (p+p')}{2N}\star j^\text{A}+\frac{iN}{2\pi}\bar a_1\wedge d\bar a_2-\frac{i(p+p')N}{4\pi}\bar a_1 \wedge d\bar a_1+\frac{i}{2\pi}\bar a_2 \wedge dA\right)\right]
~.
\fe
The first term is a decoupled $(\mathcal{Z}_N)_{-p'N}$ theory and the second term is $\tilde{\mathcal{D}}_{p+p',N}$. Together with the fact that the fusion is commutative, this gives the fusion rule
\ie\label{eq:tilde D_fusion}
\tilde{\mathcal{D}}_{p,N}\times \tilde{\mathcal{D}}_{p',N}=\tilde{\mathcal{D}}_{p',N}\times \tilde{\mathcal{D}}_{p,N}=(\mathcal{Z}_N)_{-p'N} \tilde{\mathcal{D}}_{p+p',N}=(\mathcal{Z}_N)_{-pN} \tilde{\mathcal{D}}_{p+p',N}~.
\fe
In particular, we have the non-invertible fusion
\ie
\tilde{\mathcal{D}}_{p,N}\times \tilde{\mathcal{D}}^\dagger_{p,N}=\tilde{\mathcal{D}}_{p,N}\times \tilde{\mathcal{D}}_{-p,N}=(\mathcal{Z}_N)_{-pN} \, \mathcal{C}_0\neq 1~.
\fe

As a check, we can also reproduce this fusion rule using the definition of gauging in half of the spacetime. The fusion $\tilde{\mathcal{D}}_{p,N}\times \tilde{\mathcal{D}}_{p',N}$ is described by the action
\ie
&\int_{x<0} \mathcal{L}_\text{QED}[\Psi,\bar{\Psi},A] 
    +{2\pi i(p+p') \over 2N} \oint_{x=0} \star j^\text{A}+\int_{x\geq 0} 
\left(    \mathcal{L}_\text{QED}[\Psi,\bar{\Psi},A]+ \frac{i(p+p')}{4\pi N} F \wedge F\right)\\
  &  + \int_{x\geq 0} \left(
    \frac{i}{2\pi} F\wedge b_1+\frac{iN}{2\pi}b_1\wedge dc_1
    +\frac{iN}{2\pi}b_1\wedge b_2-\frac{ip'N}{4\pi}b_2\wedge b_2+\frac{iN}{2\pi}b_2\wedge dc_2\right)
    \\
    &+\int_{x\geq 0} \left(
    \frac{iN}{2\pi} b_2\wedge b_3+\frac{iN}{2\pi}b_3\wedge dc_3
    +\frac{iN}{2\pi}b_3\wedge b_4-\frac{ipN}{4\pi}b_4\wedge b_4+\frac{iN}{2\pi}b_4\wedge dc_4\right)~.
\fe
Let $\bar b_1\equiv b_1$, $\bar b_2\equiv b_2$, $\tilde b_1\equiv b_3-pb_4$ ,$\tilde b_2\equiv (b_2+b_4)$, $\bar c_1=c_1$, $\bar c_2\equiv c_2-c_4-pc_3$, $\tilde c_1\equiv c_3$ and $\tilde c_2\equiv c_4+pc_3$. In terms of the redefined fields, the action becomes
\ie
&\int_{x<0} \mathcal{L}_\text{QED}[\Psi,\bar{\Psi},A] 
    +{2\pi i(p+p') \over 2N} \oint_{x=0} \star j^\text{A}+\int_{x\geq 0} 
\left(    \mathcal{L}_\text{QED}[\Psi,\bar{\Psi},A]+ \frac{i(p+p')}{4\pi N} F \wedge F\right)\\
  &  + \int_{x\geq 0} \left(
    \frac{i}{2\pi} F\wedge \bar b_1+\frac{iN}{2\pi}\bar b_1\wedge d\bar c_1
    +\frac{iN}{2\pi}\bar b_1\wedge \bar b_2-\frac{i(p+p')N}{4\pi}\bar b_2\wedge \bar b_2+\frac{iN}{2\pi}\bar b_2\wedge d\bar  c_2\right)
    \\
    &+\int_{x\geq 0} \left(
    \frac{iN}{2\pi} \tilde b_1\wedge\tilde b_2+\frac{iN}{2\pi}\tilde b_1\wedge d\tilde c_1
    +\frac{ipN}{4\pi}\tilde b_2\wedge \tilde b_2+\frac{iN}{2\pi}\tilde b_2\wedge d\tilde c_2\right)~.
\fe
The two lines describe the non-invertible defect $\tilde{\mathcal{D}}_{p+p',N}$ and the third line gives a decoupled $(\mathcal{Z}_N)_{-pN}$ theory at $x=0$. This reproduces the fusion rule \eqref{eq:tilde D_fusion}.

\bibliographystyle{JHEP}
\bibliography{ref}

\end{document}